\documentclass[12pt]{article}

\usepackage{graphicx}
\usepackage{amsmath}
\usepackage{amssymb}
\usepackage{amsthm}
\usepackage{pstricks}
\usepackage{color}

\renewcommand{\thefootnote}{\fnsymbol{footnote}}

\begin{document}

\begin{flushright}
CYCU-HEP-13-11 \\
KUNS-2468
\end{flushright}

\vspace{4ex}

\begin{center}

{\LARGE\bf Three-generation Asymmetric Orbifold Models from Heterotic String Theory}

\vskip 1.4cm

{\large  
Florian Beye$^{1}$\footnote{Electronic address: fbeye@eken.phys.nagoya-u.ac.jp},
Tatsuo Kobayashi$^{2}$\footnote{Electronic address: kobayash@gauge.scphys.kyoto-u.ac.jp}
and
Shogo Kuwakino$^{3}$\footnote{Electronic address: kuwakino@cycu.edu.tw}
}
\\
\vskip 1.0cm
{\it $^1$Department of Physics, Nagoya University, Furo-cho, Chikusa-ku, Nagoya 464-8602, Japan} \\
{\it $^2$Department of Physics, Kyoto University, Kitashirakawa, Kyoto 606-8502, Japan} \\
{\it $^3$Department of Physics, Chung-Yuan Christian University, 200, Chung-Pei Rd. Chung-Li,320, Taiwan} \\

\vskip 3pt
\vskip 1.5cm

\begin{abstract}
Using ${\bf Z}_3$ asymmetric orbifolds in heterotic string theory, we construct $\mathcal{N} =1$ SUSY three-generation models with the standard model gauge group $SU(3)_{{\rm C}} \times SU(2)_{{\rm L}} \times U(1)_{{\rm Y}}$ and the left-right symmetric group $SU(3)_{{\rm C}} \times SU(2)_{{\rm L}} \times SU(2)_{{\rm R}} \times U(1)_{{\rm B}-{\rm L}}$. One of the models possesses a gauge flavor symmetry for the ${\bf Z}_3$ twisted matter. 
\end{abstract}

\end{center}

\newpage

\setcounter{footnote}{0}
\renewcommand{\thefootnote}{\arabic{footnote}}

\section{Introduction}

String theory is a candidate for a unified theory of the four fundamental forces including quantum gravity. One of the main characteristic features of the standard model of particle physics is the three-generation chiral structure of quarks and leptons, and there have been many attempts to construct four-dimensional models with three generations by string compactification.

Heterotic string model building is one of the successful methods for a stringy realization of particle physics models. Especially, in the framework of heterotic ${\bf Z}_N$ or ${\bf Z}_N \times {\bf Z}_M$ orbifold compactifications \cite{orbifold}, embedding a supersymmetric standard model or a higher dimensional grand unified theory into heterotic string theory is considered \cite{heteroph,Katsuki:1989bf,Kobayashi:2004ud,Buchmuller,Kim,Lebedev:2006kn,Blaszczyk:2009in,Nibbelink:2013lua} (also see a review \cite{Nilles:2008gq}). Since strings on an orbifold can be described by a solvable world-sheet conformal field theory \cite{Hamidi:1986vh}, it is possible to calculate Yukawa couplings and selection rules \cite{Burwick:1990tu,Choi:2007nb,Kobayashi:2011cw,Bizet:2013gf}. Furthermore, the geometrical structure of orbifold fixed points can be an origin of a discrete symmetry \cite{Kobayashi:2006wq,Ko:2007dz}, which may lead to a hierarchical structure of masses/mixings of quarks and leptons.

Asymmetric orbifold compactifications \cite{Narain:1986qm} can be considered as an extension of symmetric orbifolds, in which orbifold actions for left- and right-movers are generalized to be independent of each other, without destroying modular invariance of closed string theory. We may expect that the generalization of the orbifold action will give us the possibility to construct a large number of four-dimensional string models, among which we can try to find phenomenologically viable models. However, in asymmetric orbifold constructions, model building for SUSY standard models or other GUT extended models with three matter generations has not been investigated thoroughly\footnote{In free-fermionic string constructions, which are related to ${\bf Z}_2 \times {\bf Z}_2$ (a)symmetric orbifolds by fermionization, the first SUSY standard-like models and left-right symmetric models with three generations were found in \cite{Faraggi:1989ka, Cleaver:2000ds}.}.

In \cite{Beye:2013moa}, a systematic approach for the construction of four-dimensional string models using ${\bf Z}_3$ asymmetric orbifolds, which are the simplest orbifolds capable of realizing $\mathcal{N}=1$ SUSY in four dimensions, is shown\footnote{In \cite{Ibanez:1987pj}, ${\bf Z}_3$ asymmetric orbifold models with one Wilson line were studied.}. First, one specifies a $(22,6)$-dimensional Narain lattice \cite{Narain:1985jj} and then applies the asymmetric orbifold action. In \cite{Beye:2013moa}, Narain lattices are constructed from 24-dimensional Niemeier lattices \cite{LeechNiemeier} by the lattice engineering technique\footnote{In \cite{Ito:2010df}, the lattice engineering technique is applied to a GUT model building.} \cite{Lerche:1988np}. Using this method, 106 types of (22,6)-dimensional Narain lattices with a right-moving non-Abalian factor are constructed. Group breaking patterns due to a ${\bf Z}_N$ shift action are also analyzed by extending the argument of breaking patterns of the $E_8$ group \cite{Katsuki:1989kd}. Furthermore, possible gauge group patterns of ${\bf Z}_3$ asymmetric orbifold models are analyzed.

The aim of this paper is to apply the asymmetric orbifolding procedure to the construction of three-generation models with the standard model group $SU(3)_{{\rm C}} \times SU(2)_{{\rm L}} \times U(1)_{{\rm Y}}$ and other groups. In the next section, we review asymmetric orbifold model building. In section 3 and 4 we construct ${\bf Z}_3$ asymmetric orbifold models with three generations. Section 5 is devoted to conclusions. In appendix A we show another three-generation model example.

\section{Heterotic asymmetric orbifold construction}

A starting point for the asymmetric orbifold construction is a heterotic string theory compactified on some (22,6)-dimensional Narain lattice. The corresponding world-sheet theory splits into left-moving and right-moving degrees of freedom. Besides the ghost fields, there are 26 left-moving bosons $X_{{\rm L}}$ as well as ten right-moving boson-fermion pairs $(X_{{\rm R}},\Psi_{{\rm R}})$. The momentum modes $p=( p_{{\rm L}}, p_{{\rm R}})$ associated with the internal dimensions $X_{{\rm L}}^{4 \ldots 25}$ and $X_{{\rm R}}^{4 \ldots 9}$ lie on the Narain lattice $\Gamma$. Modular invariance of closed string theory implies that $\Gamma$ is even ($p^2 = p_{{\rm L}}^2 - p_{{\rm R}}^2 \in 2 {\bf Z}$) and self-dual ($\Gamma = \tilde{\Gamma}$). Here, $\Gamma \equiv \sum_{i} n_i \gamma_i$ and $\tilde{\Gamma} \equiv \sum_{i} m_i \tilde{\gamma}_i$, where $\gamma_i$ and $\tilde{\gamma}_j$ are bases for the lattice $\Gamma$ and its dual lattice $\tilde{\Gamma}$, 
respectively. The bases for the lattice and its dual satisfy  $\gamma_i \cdot \tilde{\gamma}_j = \delta_{ij}$.

In ${\bf Z}_N$ orbifold constructions, one specifies an orbifold action for the internal dimensions as follows: 
\begin{align}
X_{{\rm L}} &\to \theta_{{\rm L}} X_{{\rm L}} + V_{{\rm L}}, \\
X_{{\rm R}} &\to \theta_{{\rm R}} X_{{\rm R}} + V_{{\rm R}}, \\
\Psi_{{\rm R}} &\to \theta_{{\rm R}} \Psi_{{\rm R}}.
\end{align}
This action contains a twist $\theta = ( \theta_{{\rm L}}, \theta_{{\rm R}} )$ and a shift $V = ( V_{{\rm L}}, V_{{\rm R}} )$. The twist has to be a lattice automorphism of order $N$, i.e. $\theta^N = 1$. The shift has to satisfy $N V \in \Gamma$. In asymmetric orbifolds, the left-mover twist $\theta_{{\rm L}}$ and the right-mover twist $\theta_{{\rm R}}$ can be chosen independently. For the sake of simplicity we consider only ${\bf Z}_3$ models with $\mathcal{N}=1$ SUSY and without a left-moving twist, i.e. $\theta_{{\rm L}} =1$ (also, one can set $V_{{\rm R}} = 0$). By writing $\theta_{{\rm R}} = {\rm diag} ( e^{2\pi i t_{\rm R}^1},e^{2\pi i t_{\rm R}^2},e^{2\pi i t_{\rm R}^3})$ in some complex basis one can define a right-mover twist vector  $t_{{\rm R}} = ( 0, t_{{\rm R}}^1, t_{{\rm R}}^2, t_{{\rm R}}^3 )$. When acting on fermions one has to embed the twist into the double cover of ${\rm SO}(6)$, so $t_{{\rm R}}^i$ is only defined modulo 2. If $\sum_i t_{{\rm R}}^i = 0$ and $t_{{
 \rm R}}^{1,2,3} \neq 0$ one realizes $\mathcal{N}=1$ SUSY. For our models we use the right-mover twist vector 
\begin{align} \label{RightTwist}
t_{{\rm R}} = ( 0, \frac{1}{3}, \frac{1}{3},- \frac{2}{3} ). 
\end{align} 
Now, we can fully specify a model by the following:
\begin{itemize}
\item a (22,6)-dimensional Narain lattice $\Gamma$ which contains a right-moving $\overline{E}_6$ or $\overline{A}_2^3$ lattice. These are the only lattices which allow for a $\mathcal{N}=1$ compatible $\mathbf{Z}_3$ automorphism \cite{Katsuki:1989bf,Kobayashi:1991rp}. In \cite{Beye:2013moa}, such lattices were constructed from the well known 24-dimensional even self-dual lattices by the lattice engineering method.
\item a ${\bf Z}_3$ shift vector $V = (V_{{\rm L}}, 0)$ that satisfies $3V \in \Gamma$. 
\end{itemize}

In heterotic string theory, spacetime gauge symmetry is realized by left-moving massless modes. Generally, a four-dimensional Narain model has a gauge symmetry of rank 22. An orbifold shift $V_{{\rm L}}$ breaks the original group into a subgroup of same rank. The breaking patterns can be calculated by analyzing shift vectors and extended Dynkin diagrams \cite{Beye:2013moa}.

In the case of ${\bf Z}_3$ asymmetric orbifold models with $\mathcal{N}=1$ SUSY whose twist action for the left-mover is trivial, we can check that modular invariance of the closed string theory requires the shift vector $V_{{\rm L}}$ to satisfy the condition
\begin{align} \label{modinv}
\frac{3V_{{\rm L}}^2}{2} \in {\bf Z}.
\end{align}

The massless spectrum in the untwisted sector can be read off as in the case of symmetric orbifolds. In the light-cone formalism, the right-moving modes can be described in terms of $H$-momentum. In the untwisted sector, the orbifold phases for $H$-momentum modes
\begin{align}
\vert q \rangle \in \{ \vert { \underline{\pm 1,0,0,0} } \rangle, 
\  \vert \pm \frac{1}{2}, \pm \frac{1}{2}, \pm \frac{1}{2}, \pm \frac{1}{2} \rangle_{+{{\rm even}}} \}
\end{align}
are given by $t_{{\rm R}} \cdot q$. Here, "+even" means that we should take only combinations with the even number of plus signs and the underline represents all cyclic permutations. The massless modes with non-trivial phases under the orbifold action $t_{{\rm R}}$ are given by
\begin{align} \label{HMomentumChiral}
\vert q' \rangle &\in 
\{ \vert 0, \underline{1,0,0} \rangle, \  
\vert \frac{1}{2}, \underline{\frac{1}{2},-\frac{1}{2},-\frac{1}{2}} \rangle \} : \ t_{{\rm R}} \cdot q' \sim 1/3, \\
\vert q'' \rangle &\in 
\{ \vert 0, \underline{-1,0,0} \rangle, \ 
\vert -\frac{1}{2}, \underline{-\frac{1}{2},\frac{1}{2},\frac{1}{2}} \rangle \} : \ t_{{\rm R}} \cdot q'' \sim -1/3.
\end{align}
Note that the $H$-momentum states $\vert q' \rangle$ and $\vert q'' \rangle $ are CPT conjugate to each other. We define the four-dimensional chirality as "left-handed" if the first component of fermionic modes is $1/2$, i.e. $\vert \frac{1}{2}, \underline{\frac{1}{2},-\frac{1}{2},-\frac{1}{2}} \rangle $ in $\vert q' \rangle$. For the lattice part, the orbifold action $(\theta, V)$ acts on momentum modes $p = ( p_{{\rm L}}, p_{{\rm R}} ) \in \Gamma$ and oscillator modes. The right-moving modes with $p_{{\rm R}} \neq 0$ or oscillator excitations are massive, so we do not need to consider orbifold phases for them. In the left-mover part, orbifold phases $V_{{\rm L}} \cdot p_{{\rm L}}$ arise for massless states with $p_{{\rm L}}^2 = 2$. Left-mover oscillators are not affected by the orbifold action because $\theta_L = 1$. This is the reason for the observed rank preservation. Now, when combining left-mover and right-mover phases one concludes that only massless states which satisfy the condition
\begin{align} \label{OrbifoldCondition}
p_{{\rm L}} \cdot V_{{\rm L}} - t_{{\rm R}} \cdot q \in {\bf Z}
\end{align}
remain in the untwisted spectrum after the orbifold projection.

To read off massless states in the ${\bf Z}_3$ twisted sector, we define  $I_{\theta} (\Gamma )$ as the sublattice of the original (22,6)-dimensional Narain lattice $\Gamma$ that is invariant under the ${\bf Z}_3$ twist action $\theta$. In our case where we have no twist action for the left-movers, $I_{\theta} (\Gamma )$ is a 22-dimensional left-mover lattice which is spanned by 22 basis vectors $\alpha_{i = 1 \ldots 22}$. We also define the dual lattice of the invariant sublattice $\tilde{I}_{\theta} (\Gamma ) \equiv \sum_{i=1}^{22} n_i \tilde{\alpha}_i$, where the $\tilde{\alpha}_i$ satisfy $\alpha_i \cdot \tilde{\alpha}_j = \delta_{ij}$ and $n_i$ are integers. In the $\alpha$-twisted sector ($\alpha \in \lbrace 1, 2\rbrace$), the momenta $p = (p_{\rm L}, 0)$ lie on the shifted lattice $\tilde{I}_\theta (\Gamma ) + \alpha V$. The massless left-mover modes in the $\alpha$-twisted sector can be obtained by solving the equation 
\begin{align} \label{MasslessForTwisted}
	\frac{p_L^2}{2} + \Delta c_{{\rm L}} -1 = 0.
\end{align}
Since in our case we do not consider any twist actions for the left-mover, we have $\Delta c_{{\rm L}} = 0$. For the right-moving part, massless states are described solely in terms of $H$-momentum:
\begin{align} \label{TwistedHM}
\vert q_0^{\alpha=1} \rangle & \in \left\lbrace \vert 0, \frac{1}{3}, \frac{1}{3}, \frac{1}{3} \rangle, \ 
\vert \frac{1}{2}, - \frac{1}{6}, - \frac{1}{6}, - \frac{1}{6} \rangle \right\rbrace \ \ ( \alpha =1 ), \\
\vert q_0^{\alpha=2} \rangle & \in \left\lbrace \vert 0, -\frac{1}{3}, -\frac{1}{3}, -\frac{1}{3} \rangle, \ 
\vert -\frac{1}{2},  \frac{1}{6},  \frac{1}{6},  \frac{1}{6} \rangle \right\rbrace \ \ ( \alpha =2 ).
\end{align}
These are combined with massless states from the lattice part $\vert p_0^{\alpha =1,2} \rangle$ that satisfy \eqref{MasslessForTwisted}, giving
\begin{align}
&\vert p_0^{\alpha =1} \rangle \otimes \vert q_0^{\alpha=1} \rangle \ \ ( \alpha =1 ) \\
&\vert p_0^{\alpha =2} \rangle \otimes \vert q_0^{\alpha=2} \rangle \ \ ( \alpha =2 ) \;.
\end{align}
The states from $\alpha = 1$ and $\alpha = 2$ are CPT conjugate to each other and together form a four-dimensional chiral supermultiplet. 
The degeneracy factor for the twisted sector is given by
\begin{align} \label{Degeneracy}
	D = \frac{ \prod_{ \{i \vert \eta^i \neq 0 \} } 
	\left[ 2 \sin(\pi \eta^i ) \right] }{ \sqrt{ {\rm Vol} 
	( I_{\theta}( \Gamma ) ) } },
\end{align}
where $\eta^i$ are defined as $0 \leq t^i + n^i = \eta^i < 1 $ for suitable integers $n^i$. The volume of the invariant lattice is determined as $ {\rm Vol} ( I_{\theta}( \Gamma ) ) = \det (g_{ij} ) = \det ( \alpha_i \cdot \alpha_j ) $. In this paper we consider only orbifold models with the right-mover twist \eqref{RightTwist} and a right-mover $\overline{E}_6$ or a $\overline{A}_2^3$ lattice. For these lattices one retrieves from \eqref{Degeneracy} a degeneracy of $D=3$ and $D=1$, respectively.

In \cite{Kakushadze:1997ub}, couplings of asymmetric orbifold models are calculated by rewriting a right-moving complex chiral boson $X(\overline{z})$ in terms of exponentials of boson fields $\phi$,
\begin{align}
i \partial X( \overline{z} ) = e^{ i \alpha \cdot \phi ( \overline{z} ) } + \cdots, 
\end{align}
where $\alpha$ corresponds to suitable root vectors. In order to reproduce the ${\bf Z}_N$ orbifold condition for $i \partial X $,
\begin{align}
i \partial X ( \overline{z} e^{-2 \pi i } ) = e^{ -2\pi i k/N } i \partial X ( \overline{z} )
\end{align}
for some integer $k$, the orbifold action on $\phi$ has to be a shift action:
\begin{align}
\phi ( \overline{z} e^{-2 \pi i } ) = \phi ( \overline{z} ) - 2 \pi s^{Q}.
\end{align}
In this representation, the momenta associated to $\phi$ (the Q-charges) enter the right mover mass equation. In particular, the massless states satisfy
\begin{align}
\frac{Q^2}{2} = \Delta c_{{\rm R}}  = \frac{1}{3},
\end{align}
where $Q \in \Gamma_{22,6} + \alpha s^{Q}$ for $\alpha = 1, 2$. The number of solutions to this equation corresponds to the degeneracy factor $D$. Now, let us denote the simple roots and fundamental weights of a simple Lie group $G$ as $\alpha_i^{G}$ and $\omega_i^{G}$. Then, for models from Narain lattices with $\overline{E}_6$, the right-moving twist \eqref{RightTwist} is replaced by the following shift for the internal six dimensions: 
\begin{align}
s^{Q}_{R} = ( \omega_2^{A_2}, 0,0 ).
\end{align}
Here we use the decomposition $A_2^3 \subset E_6$. For models from Narain lattices with $\overline{A}_2^3$, the corresponding shift action for each $A_2$ factor is given by
\begin{align}
s^{Q}_{R} =  \frac{\alpha_1^{A_2}}{3} .
\end{align}

\section{Three-generation $SU(3)_{{\rm C}} \times SU(2)_{{\rm L}} \times U(1)_{{\rm Y}}$ models from asymmetric orbifolds}

In this section we present a ${\bf Z}_3$ asymmetric orbifold model with the standard model gauge group $SU(3)_{{\rm C}} \times SU(2)_{{\rm L}} \times U(1)_{{\rm Y}}$, following the construction in \cite{Beye:2013moa}. The model is constructed from a Narain lattice with a right-moving $\overline{E}_6$ factor, i.e. it has a degeneracy $D=3$ in the twisted sectors. Another such model is shown in Appendix \ref{App1}.

As starting point we choose the 24-dimensional $A_3^8$ Niemeier lattice with conjugacy classes given by the generators
\begin{align}
(3, \underline{2, 0, 0, 1, 0, 1, 1}).
\end{align} 
Using the lattice engineering technique, we replace an $A_2$ factor contained in one of the $A_3$ of $A_3^8$ with a right-moving $\overline{E}_6$ factor as
\begin{align}
	&A_{3}^8 
	\xrightarrow[{\rm decompose}]{} A_{3}^7 \times A_2 \times U(1)
	\xrightarrow[{\rm replace}]{} A_{3}^7 \times \overline{E}_6 \times U(1).
	\nonumber
\end{align}
The resulting lattice is a (22,6)-dimensional Narain lattice with $A_3^7 \times U(1) \times \overline{E}_6$ whose conjugacy classes are generated by
\begin{align} 
&(0,0,0,0,0,0,0,1/3,2), \nonumber \\
&(0,1,1,2,0,0,1,1/4,0), \nonumber \\
&(1,0,1,1,2,0,0,1/4,0), \nonumber \\
&(0,1,0,1,1,2,0,1/4,0), \nonumber \\
&(0,0,1,0,1,1,2,1/4,0)
\end{align}
of $ A_3^{7} \times U(1) \times \overline{E}_6 $. Here, the normalization for the $U(1)$ is given by $2\sqrt{3}$. At this stage, the compactified four-dimensional model possesses $\mathcal{N}=4$ SUSY and a $SU(4)^7 \times U(1)$ gauge symmetry.

The gauge symmetry $A_3^7 \times U(1)$ will be broken by the ${\bf Z}_3$ left-mover shift action. Possible ${\bf Z}_3$ shift vectors $V$ have to satisfy $3V \in I_{\theta}( \Gamma )$ where $I_{\theta}( \Gamma )$ represents the 22-dimensional sublattice invariant under the twist. In our case, the conjugacy classes of $I_{\theta}( \Gamma )$   are given by the generators
\begin{align} \label{InvSublattice}
&(0,1,1,2,0,0,1,1/4),  \nonumber \\
&(1,0,1,1,2,0,0,1/4),  \nonumber \\
&(0,1,0,1,1,2,0,1/4),  \nonumber \\
&(0,0,1,0,1,1,2,1/4)
\end{align}
of $A_3^{7} \times U(1)$. In general, the shift vector $V$ is composed of shift vectors for the seven $A_3$ parts and a shift vector for $U(1)$ part. The shift vector for each $A_3$ belongs to one of the conjugacy classes $0, 1, 2$ and $3$. Now, before looking at the full broken gauge group, it is useful to consider $A_3$ group breaking patterns by ${\bf Z}_3$ shift actions in general. The breaking patterns are listed in Table \ref{Tab:A3breaking} (For the definition of $n_i'^{(k)}$, see \cite{Beye:2013moa}). In the table, breaking patterns due to shift vectors which belong to conjugacy class $3_{A_3}$ are not listed as these can be reproduced from shift vectors in $1_{A_3}$ by suitable reflections. Then, with \eqref{modinv} in mind we can specify a modular invariant shift vector $V$ by appropriately combining seven $A_3$ breaking patterns with a shift in the $U(1)$ direction.

\begin{table}[t]
\begin{center}
\begin{tabular}{|c|c|c|c|c|c|}
\hline
No. & C.C. & Group breaking & $n_i'^{(k)}$ & Shift vector ($3V$) & $(3V)^2 $ \\
\hline
\hline
 $1$ & $0$ & $A_3$ & $(0,0,0;3)$ & $0$ & $0$ \\
\hline
 $2$ & $0$ & $A_2 \times U(1)$ & $(0,1,2;0)$ & $\omega^{A_3}_2 + 2\omega^{A_3}_3$ & $6$  \\
\hline
 $3$ & $0$ & $A_1^2 \times U(1)$ & $(0,2,0;1)$ & $2 \omega^{A_3}_2 $ & $4$  \\
\hline
 $4$ & $0$ & $ A_1 \times U(1)^2$ & $(1,0,1;1)$ &$\omega^{A_3}_1 + \omega^{A_3}_3$ & $2$  \\
\hline
 $5$ & $0$ & $A_2 \times U(1)$ & $(2,1,0;0)$ &$2\omega^{A_3}_1 + \omega^{A_3}_2$ & $6$  \\
\hline
\hline
 $6$ & $1$ & $A_3$ & $(0,0,3;0)$ & $ 3\omega^{A_3}_3$ & $\frac{27}{4}$  \\
\hline
 $7$ & $1$ & $A_1 \times U(1)^2 $ & $(0,1,1;1)$ & $\omega^{A_3}_2 + \omega^{A_3}_3$ & $\frac{11}{4}$  \\
\hline
 $8$ & $1$ & $A_2 \times U(1)$ & $(1,0,0;2)$ & $\omega^{A_3}_1 $ & $\frac{3}{4}$  \\
\hline
 $9$ & $1$ & $A_2 \times U(1)$ & $(1,2,0;0)$ & $\omega^{A_3}_1 + 2\omega^{A_3}_2$ & $\frac{27}{4}$  \\
\hline
 $10$ & $1$ & $A_1^2 \times U(1)$ & $(2,0,1;0)$ & $2\omega^{A_3}_1 + \omega^{A_3}_3$ & $\frac{19}{4}$  \\
\hline
\hline
 $11$ & $2$ & $A_2 \times U(1)$ & $(0,0,2;1)$ & $ 2\omega^{A_3}_3$ & $3$  \\
\hline
 $12$ & $2$ & $A_1^2 \times U(1)$ & $(0,1,0;2)$ & $\omega^{A_3}_2 $ & $1$  \\
\hline
 $13$ & $2$ & $A_3$ & $(0,3,0;0)$ & $3 \omega^{A_3}_2 $ &  $9$  \\
\hline
 $14$ & $2$ & $A_1 \times U(1)^2$ & $(1,1,1;0)$ & $\omega^{A_3}_1 + \omega^{A_3}_2 + \omega^{A_3}_3$ & $5$  \\
\hline
 $15$ & $2$ & $A_2 \times U(1)$ & $(2,0,0;1)$ & $2\omega^{A_3}_1 $ & $3$ \\
\hline
\end{tabular}
\caption[smallcaption]{$A_3$ group breaking patterns 
and shift vectors. 
The last component of $n_i'^{(k)}$ corresponds to $n_0'^{(k)}$.}
\label{Tab:A3breaking}
\end{center}
\end{table}

Next, we consider a specific ${\bf Z}_3$ asymmetric orbifold construction from the $A_3^7 \times U(1) \times \overline{E}_6$ lattice. We take the shift vector as
\begin{align}
V &= ( \alpha^{A_3}_1 + 2 \alpha^{A_3}_2, \alpha^{A_3}_1 + 2 \alpha^{A_3}_2, - \alpha^{A_3}_1 - 2 \alpha^{A_3}_2, 
\alpha^{A_3}_3, 0, \alpha^{A_3}_3, \alpha^{A_3}_3, 0, 0 )/3,
\end{align}
which belongs to conjugacy class $(0,0,0,0,0,0,0,0)$ of $A_3^7 \times U(1) \times \overline{E}_6$, i.e. 
\begin{align}
3V \in ( \Gamma_{A_3}, \Gamma_{A_3}, \Gamma_{A_3}, \Gamma_{A_3}, \Gamma_{A_3}, \Gamma_{A_3}, \Gamma_{A_3},  2\sqrt{3} n, \Gamma_{E_6}),
\end{align}
where $\Gamma_{A_3} (\Gamma_{E_6})$ is an $A_3 (E_6)$ root lattice and $n$ is an integer. The shift does not affect one of the $A_3$ lattices and the $U(1)$ part. Note that  by Weyl reflections $\pm (\alpha^{A_3}_1 + 2 \alpha^{A_3}_2 )$ and $ \alpha^{A_3}_3 $ are connected to $\omega^{A_3}_2 + 2\omega^{A_3}_3$ (No.2 in Table \ref{Tab:A3breaking}) and $\omega^{A_3}_1 + \omega^{A_3}_3$ (No.4) respectively. We can check that the shift vector $V$ satisfies the modular invariance condition \eqref{modinv}. The shift vector breaks the original gauge symmetry $SU(4)^7 \times U(1)$ to
\begin{align}
SU(4) \times SU(3)^3 \times SU(2)^3 \times U(1)^{10}.
\end{align}

Massless states in the untwisted sector can be read off by collecting massless states in $\mathcal{N}=4$ gauge multiples of $A_3^7 \times U(1)$ that satisfy the orbifold condition \eqref{OrbifoldCondition}. These modes $p$ are written as
\begin{align}
p = ( p^{(1)}, p^{(2)}, p^{(3)}, p^{(4)}, p^{(5)}, p^{(6)}, p^{(7)}, 0, 0 ) 
\end{align}
of $A_3^7 \times U(1) \times \overline{E}_6$ with $p^{(i=1 \ldots 7)} \in \Gamma_{A_3}$ and $p^2 = 2$. It is useful to denote the shift vector for each $A_3$ part in terms of the unbroken subgroups as 
\begin{align}
\pm (\alpha^{A_3}_1 + 2 \alpha^{A_3}_2 ) = ( \pm (\alpha^{A_2}_1 + 2 \alpha^{A_2}_2 ), 0) 
\end{align}
of $A_2 \times U(1)$, and 
\begin{align}
\alpha^{A_3}_3 =(0,0,-\sqrt{2} ) 
\end{align}
of $A_1 \times U(1)^2 $. Now we can easily read off momentum modes with orbifold phase $p \cdot V \sim 1/3$ as
\begin{align}
p^{(1,2)} &\in \{ ( \omega_1^{A_2}, \frac{2}{\sqrt{3}} ),  ( \omega_1^{A_2} - \alpha_1^{A_2}, \frac{2}{\sqrt{3}} ), 
( \omega_1^{A_2} - \alpha_1^{A_2} - \alpha_2^{A_2}, \frac{2}{\sqrt{3}} ) \}, \\
p^{(3)} &\in \{ ( \omega_2^{A_2} ,-\frac{2}{\sqrt{3}} ),  ( \omega_2^{A_2} - \alpha_2^{A_2}, -\frac{2}{\sqrt{3}} ), 
( \omega_2^{A_2} - \alpha_1^{A_2} - \alpha_2^{A_2} ,-\frac{2}{\sqrt{3}} ) \}, \\
p^{(4,6,7)} &\in \{ ( \omega_1^{A_1} ,  1,-\frac{1}{\sqrt{2}} ), ( - \omega_1^{A_1} ,  1,-\frac{1}{\sqrt{2}} ),  
( \omega_1^{A_1} , - 1,-\frac{1}{\sqrt{2}} ), ( - \omega_1^{A_1} , - 1,-\frac{1}{\sqrt{2}} ),  
(0, 0, \sqrt{2} ) \},  
\end{align}
and no solution for $p^{(5)}$. These states survive the orbifold projection when combined with the right-moving $H$-momentum states \eqref{HMomentumChiral}, resulting in the following chiral supermultiplets: 
\begin{align}
 3 & \{ ( {\bf 3}, {\bf 1}, {\bf 1}, {\bf 1}, {\bf 1}, {\bf 1}, {\bf 1}  )
+ ( {\bf 1}, {\bf 3}, {\bf 1}, {\bf 1}, {\bf 1}, {\bf 1}, {\bf 1}  )
+ ( {\bf 1}, {\bf 1}, \overline{{\bf 3}}, {\bf 1}, {\bf 1}, {\bf 1}, {\bf 1}  ) \nonumber \\
&+ 2( {\bf 1}, {\bf 1}, {\bf 1}, {\bf 2}, {\bf 1}, {\bf 1}, {\bf 1}  )
+ 2( {\bf 1}, {\bf 1}, {\bf 1}, {\bf 1}, {\bf 1}, {\bf 2}, {\bf 1}  )
+ 2( {\bf 1}, {\bf 1}, {\bf 1}, {\bf 1}, {\bf 1}, {\bf 1}, {\bf 2}  ) \nonumber \\
&+ 3( {\bf 1}, {\bf 1}, {\bf 1}, {\bf 1}, {\bf 1}, {\bf 1}, {\bf 1}  )
\}
\end{align}
of $A_2^3 \times A_1 \times A_3 \times A_1^2$ (see Table \ref{Tab:Spectrum} for $U(1)$ charges). Note that the global factor 3 comes from right-moving modes \eqref{HMomentumChiral}.

Let us now consider the ${\bf Z}_3$ twisted sector. In order to read off massless states \eqref{MasslessForTwisted} we have to evaluate the dual of the invariant sublattice $\tilde{I}_{\theta}( \Gamma )$. By taking suitable linear combinations of \eqref{InvSublattice} we can alternatively generate $I_{\theta} ( \Gamma )$ by the following generators:
\begin{align}
&(1,0,0,1,1,3,2,0), \nonumber \\
&(0,1,0,2,3,3,3,0), \nonumber \\
&(0,0,1,1,3,2,1,0), \nonumber \\
&(0,0,0,3,2,3,1,1/4).
\end{align}
From this, one verifies that $I_{\theta} ( \Gamma )$ is spanned by the basis
\begin{align}
\alpha_{1,2} &= ( \alpha^{A_3}_{1,2}, 0,0,0,0,0,0,0 ), \\
\alpha_{3,4} &= ( 0,\alpha^{A_3}_{1,2}, 0,0,0,0,0,0 ), \\
\alpha_{5,6} &= ( 0,0,\alpha^{A_3}_{1,2}, 0,0,0,0,0 ), \\
\alpha_{7 \ldots 9} &= ( 0,0,0,\alpha^{A_3}_{1 \ldots 3}, 0,0,0,0 ), \\
\alpha_{10 \ldots 12} &= ( 0,0,0,0,\alpha^{A_3}_{1 \ldots 3}, 0,0,0 ), \\
\alpha_{13 \ldots 15} &= ( 0,0,0,0,0,\alpha^{A_3}_{1 \ldots 3}, 0,0 ), \\
\alpha_{16 \ldots 18} &= ( 0,0,0,0,0,0,\alpha^{A_3}_{1 \ldots 3},0 ), \\
\alpha_{19} &= ( \omega^{A_3}_1, 0,0,\omega^{A_3}_1,\omega^{A_3}_1,\omega^{A_3}_3,\omega^{A_3}_2, 0 ), \\
\alpha_{20} &= ( 0, \omega^{A_3}_1, 0,\omega^{A_3}_2,\omega^{A_3}_3,\omega^{A_3}_3,\omega^{A_3}_3,0 ), \\
\alpha_{21} &= ( 0,0, \omega^{A_3}_1, \omega^{A_3}_1,\omega^{A_3}_3,\omega^{A_3}_2,\omega^{A_3}_1,0 ), \\
\alpha_{22} &= ( 0,0,0, \omega^{A_3}_3, \omega^{A_3}_2,\omega^{A_3}_3,\omega^{A_3}_1,1/4 ).
\end{align}
Also, we can evaluate a dual basis $\tilde{\alpha}_i$ and find that $\tilde{I}_{\theta}( \Gamma )$ is given by the following conjugacy class generators
\begin{align} \label{GenForTL}
&( 1,2,1,3,0,0,0,-1/4 ), \nonumber \\
&( 1,3,3,0,3,0,0,-1/2 ), \nonumber \\
&( 3,3,2,0,0,3,0,-1/4 ), \nonumber \\
&( 2,1,3,0,0,0,1,-1/4 ),
\end{align}
with a $U(1)$ normalization of $2/\sqrt{3}$. Now, we can read off the massless spectrum in the twisted sector by solving \eqref{MasslessForTwisted}. Taking all linear combinations of \eqref{GenForTL} results in 256 conjugacy classes for the dual invariant sublattice. Among them, here we shall show only massless states that arise from the conjugacy class $( 1,2,1,3,0,0,0,-1/4 )$. Namely, for this conjugacy class, momentum modes in the $\alpha$ twisted sector belong to
\begin{align}
p + \alpha V \in 
& ( \Gamma_{A_3} + \omega^{A_3}_1 + \frac{ \alpha ( \alpha^{A_3}_1 + 2 \alpha^{A_3}_2 ) }{3}, 
\Gamma_{A_3} + \omega^{A_3}_2 + \frac{ \alpha ( \alpha^{A_3}_1 + 2 \alpha^{A_3}_2 ) }{3}, \nonumber \\
&\Gamma_{A_3} + \omega^{A_3}_1 + \frac{ \alpha ( - \alpha^{A_3}_1 - 2 \alpha^{A_3}_2 ) }{3} , 
 \Gamma_{A_3} + \omega^{A_3}_3 + \frac{ \alpha \alpha^{A_3}_3 }{3}, 
\Gamma_{A_3},  \nonumber \\
&\Gamma_{A_3} + \frac{ \alpha \alpha^{A_3}_3 }{3}, 
\Gamma_{A_3} + \frac{ \alpha \alpha^{A_3}_3 }{3}, 
\frac{2}{ \sqrt{3}} ( n -\frac{1}{4} ) )
.
\end{align}
For $\alpha=1$, we can find a solution for \eqref{MasslessForTwisted} as
\begin{align}
( 0, \frac{1}{2 \sqrt{3}}, 0, -\frac{1}{\sqrt{3}}, p_{A_2}''' , \frac{1}{2 \sqrt{3}}, 0, \frac{1}{2}, \frac{1}{3 \sqrt{2}}, 0, 0, 0, -\frac{ \sqrt{2}}{3}, 0, 0, - \frac{ \sqrt{2}}{3},  - \frac{1}{2 \sqrt{3}} ),
\end{align}
lying in the unbroken group 
\begin{align}
\left( A_2 \times U(1) \right)^3 \times A_1 \times U(1)^2 \times A_3 \times \left( A_1 \times U(1)^2 \right)^2  \times U(1),
\end{align}
where 
\begin{align}
p_{A_2}''' \in \{ \omega_2^{A_2}, \omega_2^{A_2} - \alpha_2^{A_2}, \omega_2^{A_2} - \alpha_1^{A_2} - \alpha_2^{A_2}  \}.
\end{align}
Combined with the $H$-momentum states in \eqref{TwistedHM} this leads to three chiral supermultiplets which transform under the non-Abelian group as $( {\bf 1}, {\bf 1}, \overline{{\bf 3}}, {\bf 1}, {\bf 1}, {\bf 1}, {\bf 1} )$. Here, the degeneracy "three" comes from \eqref{Degeneracy}, with
\begin{align}
D= \frac{ \sqrt{3}^3 }{\sqrt{3}} = 3 
\end{align}
in our case. We can read off all massless states from the other conjugacy classes \eqref{GenForTL} in the same way. The resulting massless spectrum of this model is listed in Table \ref{Tab:Spectrum}. The four-dimensional gauge group is $SU(3)_{{\rm C}} \times SU(2)_{{\rm L}} \times SU(2)^2 \times SU(3)^2 \times SU(4) \times U(1)^{10}$, and there are $3 \times 12$ fields in the untwisted sector and $3 \times 37$ fields in the ${\bf Z}_3$ twisted sector. Here, all fields are labeled by $f_i (i =1 \ldots 49)$. In the table, normalizations for the $U(1)^{10}$ charges $U_{1 \ldots 10}$ are taken as 
\begin{align}
U_{1,2,3} &= \frac{2}{\sqrt{3}} Q_{1,2,3}, \\
U_{4,6,8} &= Q_{4,6,8}, \\
U_{5,7,9} &= \sqrt{2} Q_{5,7,9}, \\
U_{10} &= \frac{\sqrt{3}}{2} Q_{10}. 
\end{align}
Among the ten $U(1)$ groups, a non-anomalous $U(1)_Y$ gauge symmetry is taken from a combination of four $U(1)$s,
\begin{align} \label{U1Y}
Q_Y &= \frac{1}{2 \sqrt{3}} U_1+ \frac{1}{\sqrt{2}} U_7 -\frac{1}{2} U_8 - \frac{\sqrt{3}}{2} U_{10}.
\end{align}
By choosing this combination, we can see that this model contains three standard model generations of chiral matter multiplets, and the additional fields have vector-like structure. Also this model has one anomalous $U(1)_A$ gauge symmetry that can be given by the following combination
\begin{align}
Q_A &= \frac{1}{\sqrt{3}} U_1 + \frac{1}{\sqrt{3}} U_2 -\frac{1}{\sqrt{3}} U_3 +\frac{\sqrt{2}}{3} U_5 -\frac{1}{3 \sqrt{2}} U_7-\frac{1}{3 \sqrt{2}} U_9.
\end{align}
In a four-dimensional model with an anomalous $U(1)_A$ gauge symmetry, a string loop effect will generate a Fayet-Iliopoulos $D$-term \cite{Dine:1987xk, Atick:1987gy, Dine:1987gj}.
For the anomalous $U(1)_A$ we can check that mixed anomalies satisfy the Green-Schwarz universality relation\footnote{
See e.g. \cite{Kobayashi:1996pb} and references therein.}
\begin{align}
\frac{1}{k_a} {\rm Tr}_{G_a} T(R) Q_A 
&= {\rm Tr} Q_B^2 Q_A 
= \frac{1}{3} {\rm Tr} Q_A^3
= \frac{1}{24} {\rm Tr} Q_A
=8 \pi^2 \delta_{GS} = \sqrt{\frac{2}{3}}
,
\end{align}
where $G_a$ and $k_a$ are non-Abelian groups and their Kac-Moody levels, and $2 T(R) $ is the Dynkin index of the representation $R$. $Q_B$ represents a non-anomalous $U(1)$ group of level one that is orthogonal to $U(1)_A$.

\begin{table}[h]
\begin{center}
\scriptsize
\begin{tabular}{|c|c|c|c|cccccccccc|c|c|c|}
\hline
$U/T$ & $f$ & & ${\rm Irrep.}$ & $Q_1$ & $Q_2$ & $Q_3$ & $Q_4$ & $Q_5$ & $Q_6$ & $Q_7$ & $Q_8$ & $Q_9$ & $Q_{10}$ 
& $Q_{Y}$ & $Q_{A}$ & ${\rm Deg.}$ \\ 
\hline
$U$ & $1$ &  & $({\bf 1},{\bf 1};{\bf 1},{\bf 1},{\bf 1},{\bf 1},{\bf 1})$ & $0$ & $0$ & $0$ & $0$ & $1$ & $0$ & $0$ & $0$ & $0$ & $0$
& $0$ & $\frac{2}{3}$ & $3$ \\
$U$ & $2$ &  & $({\bf 1},{\bf 1};{\bf 1},{\bf 1},{\bf 1},{\bf 1},{\bf 1})$ & $0$ & $0$ & $0$ & $0$ & $0$ & $0$ & $1$ & $0$ & $0$ & $0$
& $1$ & $-\frac{1}{3}$ & $3$ \\
$U$ & $3$ & $s^{0u}$ & $({\bf 1},{\bf 1};{\bf 1},{\bf 1},{\bf 1},{\bf 1},{\bf 1})$ & $0$ & $0$ & $0$ & $0$ & $0$ & $0$ & $0$ & $0$ & $1$ & $0$
& $0$ & $-\frac{2}{3}$ & $3$ \\
$U$ & $4$ &  & $({\bf 1},{\bf 1};{\bf 2},{\bf 1},{\bf 1},{\bf 1},{\bf 1})$ & $0$ & $0$ & $0$ & $1$ & $-\frac{1}{2}$ & $0$ & $0$ & $0$ & $0$ & $0$
& $0$ & $-\frac{2}{3}$ & $3$ \\
$U$ & $5$ &  & $({\bf 1},{\bf 1};{\bf 2},{\bf 1},{\bf 1},{\bf 1},{\bf 1})$ & $0$ & $0$ & $0$ & $-1$ & $-\frac{1}{2}$ & $0$ & $0$ & $0$ & $0$ & $0$
& $0$ & $-\frac{2}{3}$ & $3$ \\
$U$ & $6$ &  & $({\bf 1},{\bf 1};{\bf 1},{\bf 2},{\bf 1},{\bf 1},{\bf 1})$ & $0$ & $0$ & $0$ & $0$ & $0$ & $1$ & $-\frac{1}{2}$ & $0$ & $0$ & $0$
& $-\frac{1}{2}$ & $\frac{1}{6}$ & $3$ \\
$U$ & $7$ &  & $({\bf 1},{\bf 1};{\bf 1},{\bf 2},{\bf 1},{\bf 1},{\bf 1})$ & $0$ & $0$ & $0$ & $0$ & $0$ & $-1$ & $-\frac{1}{2}$ & $0$ & $0$ & $0$
& $-\frac{1}{2}$ & $\frac{1}{6}$ & $3$ \\
$U$ & $8$ &  & $({\bf 1},{\bf 1};{\bf 1},{\bf 1},{\bf 3},{\bf 1},{\bf 1})$ & $0$ & $1$ & $0$ & $0$ & $0$ & $0$ & $0$ & $0$ & $0$ & $0$
& $0$ & $\frac{2}{3}$ & $3$ \\
$U$ & $9$ &  & $({\bf 1},{\bf 1};{\bf 1},{\bf 1},{\bf 1},\overline{{\bf 3}}, {\bf 1})$ & $0$ & $0$ & $-1$ & $0$ & $0$ & $0$ & $0$ & $0$ & $0$ & $0$
& $0$ & $\frac{2}{3}$ & $3$ \\
$U$ & $10$ & $l^u$ & $({\bf 1},{\bf 2};{\bf 1},{\bf 1},{\bf 1},{\bf 1},{\bf 1})$ & $0$ & $0$ & $0$ & $0$ & $0$ & $0$ & $0$ & $1$ & $-\frac{1}{2}$ & $0$
& $-\frac{1}{2}$ & $\frac{1}{6}$ & $3$ \\
$U$ & $11$ & $\overline{l}^u$ & $({\bf 1},{\bf 2};{\bf 1},{\bf 1},{\bf 1},{\bf 1},{\bf 1})$ & $0$ & $0$ & $0$ & $0$ & $0$ & $0$ & $0$ & $-1$ & $-\frac{1}{2}$ & $0$
& $\frac{1}{2}$ & $\frac{1}{6}$ & $3$ \\
$U$ & $12$ & $\overline{d}$ & $(\overline{{\bf 3}},{\bf 1};{\bf 1},{\bf 1},{\bf 1},{\bf 1},{\bf 1})$ & $1$ & $0$ & $0$ & $0$ & $0$ & $0$ & $0$ & $0$ & $0$ & $0$
& $\frac{1}{3}$ & $\frac{2}{3}$ & $3$ \\
%
%
$T$ & $13$ &  & $({\bf 1},{\bf 1};{\bf 1},{\bf 1},{\bf 1},{\bf 1},{\bf 1})$ 
& $-\frac{1}{2}$ & $\frac{1}{4}$ & $-\frac{1}{4}$ & $0$ & $\frac{2}{3}$ & $0$ & $-\frac{1}{3}$ & $-\frac{1}{2}$ & $\frac{1}{6}$ & $- \frac{1}{3}$
& $0$ & $\frac{1}{2}$ & $3$ \\
$T$ & $14$ &  & $({\bf 1},{\bf 1};{\bf 1},{\bf 1},{\bf 1},{\bf 1},{\bf 1})$
 & $-\frac{1}{2}$ & $\frac{1}{4}$ & $-\frac{1}{4}$ & $0$ & $-\frac{1}{3}$ & $0$ & $\frac{2}{3}$ & $-\frac{1}{2}$ & $\frac{1}{6}$ & $- \frac{1}{3}$ 
& $1$ & $-\frac{1}{2}$ & $3$ \\
$T$ & $15$ &  & $({\bf 1},{\bf 1};{\bf 1},{\bf 1},{\bf 1},{\bf 1},{\bf 1})$ 
& $-\frac{1}{2}$ & $\frac{1}{4}$ & $-\frac{1}{4}$ & $0$ & $-\frac{1}{3}$ & $0$ & $-\frac{1}{3}$ & $-\frac{1}{2}$ & $\frac{1}{6}$ & $1$
& $-1$ & $-\frac{1}{6}$ & $3$ \\
$T$ & $16$ &  & $({\bf 1},{\bf 1};{\bf 1},{\bf 1},{\bf 1},{\bf 1},{\bf 1})$ 
& $-\frac{1}{2}$ & $-\frac{1}{2}$ & $\frac{1}{2}$ & $\frac{1}{2}$ & $\frac{1}{6}$ & $\frac{1}{2}$ & $\frac{1}{6}$ & $-\frac{1}{2}$ & $\frac{1}{6}$ & $\frac{1}{3}$
& $0$ & $-1$ & $3$ \\
$T$ & $17$ &  & $({\bf 1},{\bf 1};{\bf 1},{\bf 1},{\bf 1},{\bf 1},{\bf 1})$ 
& $\frac{1}{4}$ & $\frac{1}{4}$ & $\frac{1}{2}$ & $0$ & $\frac{2}{3}$ & $-\frac{1}{2}$ & $\frac{1}{6}$ & $0$ & $-\frac{1}{3}$ & $\frac{1}{3}$
& $0$ & $\frac{1}{2}$ & $3$ \\
$T$ & $18$ &  & $({\bf 1},{\bf 1};{\bf 1},{\bf 1},{\bf 1},{\bf 1},{\bf 1})$ 
& $\frac{1}{4}$ & $\frac{1}{4}$ & $\frac{1}{2}$ & $0$ & $-\frac{1}{3}$ & $-\frac{1}{2}$ & $\frac{1}{6}$ & $0$ & $-\frac{1}{3}$ & $-1$
& $1$ & $-\frac{1}{6}$ & $3$ \\
$T$ & $19$ &  & $({\bf 1},{\bf 1};{\bf 1},{\bf 1},{\bf 1},{\bf 1},{\bf 1})$ 
& $\frac{1}{4}$ & $-\frac{1}{2}$ & $-\frac{1}{4}$ & $0$ & $\frac{2}{3}$ & $\frac{1}{2}$ & $\frac{1}{6}$ & $\frac{1}{2}$ & $\frac{1}{6}$ & $0$
& $0$ & $\frac{1}{3}$ & $3$ \\
$T$ & $20$ & $s^0_1$ & $({\bf 1},{\bf 1};{\bf 1},{\bf 1},{\bf 1},{\bf 1},{\bf 1})$ 
& $-\frac{1}{2}$ & $-\frac{1}{2}$ & $\frac{1}{2}$ & $-\frac{1}{2}$ & $\frac{1}{6}$ & $-\frac{1}{2}$ & $\frac{1}{6}$ & $\frac{1}{2}$ & $\frac{1}{6}$ & $- \frac{1}{3}$
& $0$ & $-1$ & $3$ \\
$T$ & $21$ & $s^0_2$ & $({\bf 1},{\bf 1};{\bf 1},{\bf 1},{\bf 1},{\bf 1},{\bf 1})$ 
& $\frac{1}{4}$ & $\frac{1}{4}$ & $\frac{1}{2}$ & $0$ & $-\frac{1}{3}$ & $-\frac{1}{2}$ & $\frac{1}{6}$ & $0$ & $\frac{2}{3}$ & $\frac{1}{3}$
& $0$ & $-\frac{1}{2}$ & $3$ \\
$T$ & $22$ &  & $({\bf 1},{\bf 1};{\bf 2},{\bf 1},{\bf 1},{\bf 1},{\bf 1})$ 
& $\frac{1}{4}$ & $-\frac{1}{2}$ & $-\frac{1}{4}$ & $0$ & $\frac{1}{6}$ & $-\frac{1}{2}$ & $\frac{1}{6}$ & $-\frac{1}{2}$ & $\frac{1}{6}$ & $\frac{2}{3}$
& $0$ & $0$ & $3$ \\
%
%
$T$ & $23$ &  & $({\bf 1},{\bf 1};{\bf 2},{\bf 1},{\bf 1},{\bf 1},{\bf 1})$ 
& $\frac{1}{4}$ & $-\frac{1}{2}$ & $-\frac{1}{4}$ & $0$ & $\frac{1}{6}$ & $-\frac{1}{2}$ & $\frac{1}{6}$ & $-\frac{1}{2}$ & $\frac{1}{6}$ & $- \frac{2}{3}$
& $1$ & $0$ & $3$ \\
$T$ & $24$ &  & $({\bf 1},{\bf 1};{\bf 1},{\bf 2},{\bf 1},{\bf 1},{\bf 1})$ 
& $\frac{1}{4}$ & $\frac{1}{4}$ & $\frac{1}{2}$ & $0$ & $-\frac{1}{3}$ & $\frac{1}{2}$ & $-\frac{1}{3}$ & $0$ & $-\frac{1}{3}$ & $ \frac{1}{3}$
& $-\frac{1}{2}$ & $0$ & $3$ \\
$T$ & $25$ &  & $({\bf 1},{\bf 1};{\bf 1},{\bf 2},{\bf 1},{\bf 1},{\bf 1})$ 
& $\frac{1}{4}$ & $-\frac{1}{2}$ & $-\frac{1}{4}$ & $0$ & $-\frac{1}{3}$ & $-\frac{1}{2}$ & $-\frac{1}{3}$ & $\frac{1}{2}$ & $\frac{1}{6}$ & $0$
& $-\frac{1}{2}$ & $-\frac{1}{6}$ & $3$ \\
$T$ & $26$ &  & $({\bf 1},{\bf 1};{\bf 1},{\bf 1},{\bf 3},{\bf 1},{\bf 1})$ 
& $\frac{1}{4}$ & $-\frac{1}{4}$ & $\frac{1}{2}$ & $-\frac{1}{2}$ & $\frac{1}{6}$ & $0$ & $-\frac{1}{3}$ & $-\frac{1}{2}$ & $\frac{1}{6}$ & $0$
& $0$ & $-\frac{1}{6}$ & $3$ \\
$T$ & $27$ &  & $({\bf 1},{\bf 1};{\bf 1},{\bf 1},{\bf 3},{\bf 1},{\bf 1})$
 & $-\frac{1}{2}$ & $-\frac{1}{4}$ & $-\frac{1}{4}$ & $\frac{1}{2}$ & $\frac{1}{6}$ & $-\frac{1}{2}$ & $\frac{1}{6}$ & $0$ & $-\frac{1}{3}$ & $0$
& $0$ & $-\frac{1}{6}$ & $3$ \\
$T$ & $28$ &  & $({\bf 1},{\bf 1};{\bf 1},{\bf 1},{\bf 3},{\bf 1},{\bf 1})$ 
& $\frac{1}{4}$ & $\frac{1}{2}$ & $-\frac{1}{4}$ & $0$ & $-\frac{1}{3}$ & $\frac{1}{2}$ & $\frac{1}{6}$ & $\frac{1}{2}$ & $\frac{1}{6}$ & $0$
& $0$ & $\frac{1}{3}$ & $3$ \\
$T$ & $29$ &  & $({\bf 1},{\bf 1};{\bf 1},{\bf 1},{\bf 1},{\bf 3},{\bf 1})$ 
& $\frac{1}{4}$ & $\frac{1}{4}$ & $0$ & $-\frac{1}{2}$ & $\frac{1}{6}$ & $0$ & $-\frac{1}{3}$ & $\frac{1}{2}$ & $\frac{1}{6}$ & $ \frac{2}{3}$
& $-1$ & $\frac{1}{2}$ & $3$ \\
$T$ & $30$ &  & $({\bf 1},{\bf 1};{\bf 1},{\bf 1},{\bf 1},{\bf 3},{\bf 1})$ 
& $\frac{1}{4}$ & $\frac{1}{4}$ & $0$ & $-\frac{1}{2}$ & $\frac{1}{6}$ & $0$ & $-\frac{1}{3}$ & $\frac{1}{2}$ & $\frac{1}{6}$ & $- \frac{2}{3}$
& $0$ & $\frac{1}{2}$ & $3$ \\
$T$ & $31$ &  & $({\bf 1},{\bf 1};{\bf 1},{\bf 1},{\bf 1},\overline{{\bf 3}},{\bf 1})$ 
& $\frac{1}{4}$ & $\frac{1}{4}$ & $-\frac{1}{2}$ & $0$ & $-\frac{1}{3}$ & $-\frac{1}{2}$ & $\frac{1}{6}$ & $0$ & $-\frac{1}{3}$ & $ \frac{1}{3}$
& $0$ & $\frac{1}{2}$ & $3$ \\
$T$ & $32$ &  & $({\bf 1},{\bf 1};{\bf 1},{\bf 1},{\bf 1},\overline{{\bf 3}},{\bf 1})$ 
& $-\frac{1}{2}$ & $\frac{1}{4}$ & $\frac{1}{4}$ & $-\frac{1}{2}$ & $\frac{1}{6}$ & $\frac{1}{2}$ & $\frac{1}{6}$ & $0$ & $-\frac{1}{3}$ & $0$
& $0$ & $-\frac{1}{6}$ & $3$ \\
%
%
$T$ & $33$ &  & $({\bf 1},{\bf 1};{\bf 1},{\bf 1},{\bf 1},\overline{{\bf 3}},{\bf 1})$ 
& $\frac{1}{4}$ & $-\frac{1}{2}$ & $\frac{1}{4}$ & $\frac{1}{2}$ & $\frac{1}{6}$ & $0$ & $-\frac{1}{3}$ & $0$ & $-\frac{1}{3}$ & $- \frac{1}{3}$
& $0$ & $0$ & $3$ \\
$T$ & $34$ &  & $({\bf 1},{\bf 1};{\bf 1},{\bf 1},{\bf 1},{\bf 1},{\bf 4})$ 
& $\frac{1}{4}$ & $\frac{1}{4}$ & $-\frac{1}{4}$ & $-\frac{1}{2}$ & $\frac{1}{6}$ & $\frac{1}{2}$ & $\frac{1}{6}$ & $-\frac{1}{2}$ & $\frac{1}{6}$ & $ \frac{1}{3}$
& $\frac{1}{4}$ & $\frac{1}{2}$ & $3$ \\
$T$ & $35$ &  & $({\bf 1},{\bf 1};{\bf 1},{\bf 1},{\bf 1},{\bf 1},{\bf 4})$ 
& $\frac{1}{4}$ & $\frac{1}{4}$ & $-\frac{1}{4}$ & $\frac{1}{2}$ & $\frac{1}{6}$ & $-\frac{1}{2}$ & $\frac{1}{6}$ & $\frac{1}{2}$ & $\frac{1}{6}$ & $- \frac{1}{3}$
& $\frac{1}{4}$ & $\frac{1}{2}$ & $3$ \\
$T$ & $36$ &  & $({\bf 1},{\bf 1};{\bf 2},{\bf 1},{\bf 1},{\bf 1},\overline{{\bf 4}})$ 
& $\frac{1}{4}$ & $\frac{1}{4}$ & $-\frac{1}{4}$ & $0$ & $\frac{1}{6}$ & $0$ & $-\frac{1}{3}$ & $0$ & $-\frac{1}{3}$ & $0$
& $-\frac{1}{4}$ & $\frac{5}{6}$ & $3$ \\
$T$ & $37$ &  & $({\bf 1},{\bf 1};{\bf 2},{\bf 2},{\bf 1},{\bf 1},{\bf 1})$ 
& $-\frac{1}{2}$ & $\frac{1}{4}$ & $-\frac{1}{4}$ & $0$ & $\frac{1}{6}$ & $0$ & $\frac{1}{6}$ & $\frac{1}{2}$ & $\frac{1}{6}$ & $ \frac{1}{3}$
& $-\frac{1}{2}$ & $0$ & $3$ \\
$T$ & $38$ &  & $({\bf 1},{\bf 1};{\bf 1},{\bf 2},\overline{{\bf 3}},{\bf 1},{\bf 1})$ 
& $\frac{1}{4}$ & $0$ & $-\frac{1}{4}$ & $-\frac{1}{2}$ & $\frac{1}{6}$ & $0$ & $\frac{1}{6}$ & $0$ & $-\frac{1}{3}$ & $- \frac{1}{3}$
& $\frac{1}{2}$ & $\frac{1}{2}$ & $3$ \\
$T$ & $39$ &  & $({\bf 1},{\bf 1};{\bf 1},{\bf 2},{\bf 1},{\bf 3},{\bf 1})$ 
& $\frac{1}{4}$ & $\frac{1}{4}$ & $0$ & $\frac{1}{2}$ & $\frac{1}{6}$ & $0$ & $\frac{1}{6}$ & $-\frac{1}{2}$ & $\frac{1}{6}$ & $0$
& $\frac{1}{2}$ & $\frac{1}{3}$ & $3$ \\
%
%
$T$ & $40$ & $c_1$ & $({\bf 3},{\bf 1};{\bf 1},{\bf 1},{\bf 1},{\bf 1},{\bf 1})$ 
& $0$ & $\frac{1}{4}$ & $-\frac{1}{4}$ & $\frac{1}{2}$ & $\frac{1}{6}$ & $\frac{1}{2}$ & $\frac{1}{6}$ & $0$ & $-\frac{1}{3}$ & $ \frac{2}{3}$
& $-\frac{1}{3}$ & $\frac{1}{2}$ & $3$ \\
$T$ & $41$ & $c_2$ & $({\bf 3},{\bf 1};{\bf 1},{\bf 1},{\bf 1},{\bf 1},{\bf 1})$ 
& $0$ & $\frac{1}{4}$ & $-\frac{1}{4}$ & $\frac{1}{2}$ & $\frac{1}{6}$ & $\frac{1}{2}$ & $\frac{1}{6}$ & $0$ & $-\frac{1}{3}$ & $- \frac{2}{3}$
& $\frac{2}{3}$ & $\frac{1}{2}$ & $3$ \\
$T$ & $42$ & $\overline{c}_1$ & $(\overline{{\bf 3}},{\bf 1};{\bf 1},{\bf 1},{\bf 1},{\bf 1},{\bf 1})$ 
& $\frac{1}{2}$ & $\frac{1}{4}$ & $-\frac{1}{4}$ & $0$ & $-\frac{1}{3}$ & $0$ & $-\frac{1}{3}$ & $-\frac{1}{2}$ & $\frac{1}{6}$ & $- \frac{1}{3}$
& $\frac{1}{3}$ & $\frac{1}{2}$ & $3$ \\
$T$ & $43$ & $\overline{c}_2$ & $(\overline{{\bf 3}},{\bf 1};{\bf 1},{\bf 1},{\bf 1},{\bf 1},{\bf 1})$ 
& $-\frac{1}{4}$ & $-\frac{1}{2}$ & $-\frac{1}{4}$ & $-\frac{1}{2}$ & $\frac{1}{6}$ & $0$ & $-\frac{1}{3}$ & $0$ & $-\frac{1}{3}$ & $ \frac{1}{3}$
& $-\frac{2}{3}$ & $0$ & $3$ \\
$T$ & $44$ &  & $({\bf 1},{\bf 2};{\bf 1},{\bf 1},{\bf 1},{\bf 1},{\bf 1})$ 
& $-\frac{1}{2}$ & $\frac{1}{4}$ & $-\frac{1}{4}$ & $0$ & $-\frac{1}{3}$ & $0$ & $-\frac{1}{3}$ & $\frac{1}{2}$ & $-\frac{1}{3}$ & $- \frac{1}{3}$
& $-\frac{1}{2}$ & $0$ & $3$ \\
$T$ & $45$ &  & $({\bf 1},{\bf 2};{\bf 2},{\bf 1},{\bf 1},{\bf 1},{\bf 1})$ 
& $\frac{1}{4}$ & $\frac{1}{4}$ & $\frac{1}{2}$ & $0$ & $\frac{1}{6}$ & $\frac{1}{2}$ & $\frac{1}{6}$ & $0$ & $\frac{1}{6}$ & $- \frac{1}{3}$
& $\frac{1}{2}$ & $0$ & $3$ \\
$T$ & $46$ &  & $({\bf 1},{\bf 2};{\bf 1},{\bf 1},\overline{{\bf 3}},{\bf 1},{\bf 1})$ 
& $\frac{1}{4}$ & $0$ & $-\frac{1}{4}$ & $\frac{1}{2}$ & $\frac{1}{6}$ & $0$ & $-\frac{1}{3}$ & $0$ & $\frac{1}{6}$ & $ \frac{1}{3}$
& $-\frac{1}{2}$ & $\frac{1}{2}$ & $3$ \\
$T$ & $47$ & $q$ & $({\bf 3},{\bf 2};{\bf 1},{\bf 1},{\bf 1},{\bf 1},{\bf 1})$ 
& $0$ & $\frac{1}{4}$ & $-\frac{1}{4}$ & $-\frac{1}{2}$ & $\frac{1}{6}$ & $-\frac{1}{2}$ & $\frac{1}{6}$ & $0$ & $\frac{1}{6}$ & $0$
& $\frac{1}{6}$ & $\frac{1}{3}$ & $3$ \\
$T$ & $48$ & $\overline{u}$ & $(\overline{{\bf 3}},{\bf 1};{\bf 1},{\bf 1},{\bf 1},{\bf 1},{\bf 1})$ 
& $-\frac{1}{4}$ & $\frac{1}{4}$ & $\frac{1}{2}$ & $\frac{1}{2}$ & $\frac{1}{6}$ & $0$ & $-\frac{1}{3}$ & $\frac{1}{2}$ & $\frac{1}{6}$ & $0$
& $-\frac{2}{3}$ & $-\frac{1}{6}$ & $3$ \\
$T$ & $49$ & $h_u$ & $({\bf 1},{\bf 2};{\bf 1},{\bf 1},{\bf 1},{\bf 1},{\bf 1})$ 
& $\frac{1}{4}$ & $-\frac{1}{2}$ & $-\frac{1}{4}$ & $0$ & $-\frac{1}{3}$ & $\frac{1}{2}$ & $\frac{1}{6}$ & $-\frac{1}{2}$ & $-\frac{1}{3}$ & $0$
& $\frac{1}{2}$ & $-\frac{1}{6}$ & $3$ \\
\hline
\end{tabular}
\caption[smallcaption]{Massless spectrum of three-generation $SU(3)_{{\rm C}} \times SU(2)_{{\rm L}} \times U(1)_{{\rm Y}}$ model. Representations under the non-Abelian group $SU(3)_{{\rm C}} \times SU(2)_{{\rm L}} \times SU(2)^2 \times SU(3)^2 \times SU(4)$ and $U(1)$ charges are listed. U and T mean the untwisted and twisted sector respectively. Note that all fields have degeneracy 3. The gravity and gauge supermultiplets are omitted.}
\label{Tab:Spectrum}
\end{center}
\end{table}

By the $U(1)_{{\rm Y}}$ charge assignment \eqref{U1Y} we can see that this model has net three standard model generations, and the other extra fields are vector-like to each other. In this model, three-generations of right-handed down-type quarks $\overline{d}$ come from the untwisted sector (field $f_{12}$), and further three generations of quark doublets $q$, right-handed down-type quarks $\overline{u}$ and up-type Higgs fields $h_u$ come from the ${\bf Z}_3$ twisted sector\footnote{It is also possible to realize the up-type quark and up-type Yukawa coupling from the untwisted sector by choosing other Narain lattices and shift vectors. } (fields $f_{47}, f_{48}$ and $f_{49}$). We find two pairs of $SU(3)_{{\rm C}}$ color exotics and some extra $SU(2)_{{\rm L}}$ doublets. There are also $3\times 12$ singlets of the non-Abelian group $SU(3)_{{\rm C}} \times SU(2)_{{\rm L}} \times SU(2)^2 \times SU(3)^2 \times SU(4)$, and all other $SU(3)_{{\rm C}} \times SU(2)_{{\rm L}}$ singlets are
  non-trivially charged under the additional non-Abelian group $SU(2)^2 \times SU(3)^2 \times SU(4)$.

In this paper we do not consider explicitly all of the terms in the superpotential of this model and we do not perform detailed analysis of the VEV structure.  That is our future task. In the following, we will comment on decoupling of some exotic fields and Yukawa couplings of this model. Regarding the color exotic fields, $c_1, c_2, \overline{c}_1$ and $\overline{c}_2$ have a three point coupling with singlets as 
\begin{align}
s_1^0 c_1 \overline{c}_1 , \ \ s_2^0 c_2 \overline{c}_2,
\end{align}
so they are expected to decouple from the low energy theory if the singlets $s_1^0$ and $s_2^0$ get VEVs. Similar thing can happen for $l^u$ and $\overline{l}^u$ since there is a three point coupling 
\begin{align}
s^{0u} l^u \overline{l}^u.
\end{align}
Fields $f_{44}, f_{45}$ and $f_{46}$ contain four $({\bf 1}, {\bf 2})_{-1/2}$ fields and two $({\bf 1}, {\bf 2})_{1/2}$ fields after breaking additional $A_3$ and $A_2$ groups. Then, we can expect that net two $({\bf 1}, {\bf 2})_{-1/2}$ remain massless fields, and these fields can be identified as the lepton doublet and down-type Higgs fields. We can see that the other exotic fields have vector-like structure.

By analyzing three point couplings allowed by Q-charge invariance, relevant couplings for quarks and Higgses are given by
\begin{align}
y_{123} q_1 h_{u2} \overline{u}_3 + y_{312} q_3 h_{u1} \overline{u}_2 + y_{231} q_2 h_{u3} \overline{u}_1 + 
y_{132} q_1 h_{u3} \overline{u}_2 + y_{213} q_2 h_{u1} \overline{u}_3 + y_{321} q_3 h_{u2} \overline{u}_1,
\end{align}
because the Q-charges are
\begin{align}
	Q(X_1) &= ( \omega_2^{A_2}, 0,0 ),\\
	Q(X_2) &= ( \omega_2^{A_2} - \alpha_2^{A_2}, 0,0 ),\\
	Q(X_3) &= ( \omega_2^{A_2} - \alpha_2^{A_2} - \alpha_1^{A_2}, 0,0 )
\end{align} for $X \in \{ q, h_u, \overline{u}\}$. Since in asymmetric orbifolds the starting point is a torus compactification at self-dual radius, string world-sheet instanton effects can be neglected. So, each coefficient $y_{123}$ {etc.} for the three point couplings is expected to be of $\mathcal{O}(1)$. From above couplings it turns out that the mass of the second generation quark will be of the same order as the top quark mass.

\section{Three-generation $SU(3)_{{\rm C}} \times SU(2)_{{\rm L}} \times SU(2)_{{\rm R}} \times U(1)_{{\rm B}-{\rm L}}$ models from asymmetric orbifolds}

In this section we construct a ${\bf Z}_3$ asymmetric orbifold model with the left-right symmetric group  $SU(3)_{{\rm C}} \times SU(2)_{{\rm L}} \times SU(2)_{{\rm R}} \times U(1)_{{\rm B}-{\rm L}}$. We use a Narain lattice with $\overline{A}_2^3$ as our starting point, so the resulting model has a degeneracy factor $D=1$.

We start from four-dimensional heterotic string theory compactified on a (20,4)-dimensional Narain lattice $A_1^2 \times A_4^4 \times U(1)^2 \times \overline{A}_2^2 $ and a (2,2)-dimensional $A_2 \times \overline{A}_2$ lattice. Using the lattice engineering technique, the (20,4)-dimensional lattice is made from the 24-dimensional $A_4^6$ Niemeier lattice as
\begin{align}
	&A_{4}^6 
	\xrightarrow[{\rm decompose}]{} \left( A_2 \times A_1 \times U(1) \right)^2 \times A_{4}^4 
	\xrightarrow[{\rm replace}]{} A_1^2 \times A_{4}^4 \times U(1)^2 \times \overline{A}_2^2 .
	\nonumber
\end{align}
Here, the left-moving $A_2^2$ factor in $A_4^2$ is replaced by the right-moving $\overline{A}_2^2$ factor. Corresponding conjugacy class generators are given by
\begin{align} 
&(1,0,0,0,0,0,1/6,0,1,2), \nonumber \\
&(0,1,1,4,4,1,1/15,1/6,0,1), \nonumber \\
&(0,0,4,1,0,1,1/15,-1/15,2,0), \nonumber \\
&(0,0,4,4,1,0,1/15,1/15,0,1)
\end{align}
of $ A_1^2 \times A_4^{4} \times U(1)^2 \times \overline{A}_2^2 $. Here, the normalizations for the two $U(1)$s are taken as $\sqrt{30}$. At this stage, the compactified four-dimensional model possesses $\mathcal{N}=4$ SUSY and $SU(5)^4 \times SU(3) \times SU(2)^2 \times U(1)^2$ gauge symmetry.

Gauge symmetry breaking patterns for this model can be analyzed as in the previous section. Breaking patterns for $A_1$, $A_2$ and $A_4$ groups are listed in Table \ref{Tab:A1breaking}, \ref{Tab:A2breaking} and \ref{Tab:A4breaking}.

\begin{table}[ht]
\begin{center}
\begin{tabular}{|c|c|c|c|c|c|}
\hline
No. & C.C. & Group breaking & $n_i'^{(k)}$ & Shift vector ($3V$) & $(3V)^2 $ \\
\hline
\hline
$1$ & $0$ & ${{A_1}}$ & $(0;3)$ & $ 0 $ & $0$ \\
\hline
$2$ & $0$ & ${{U(1)}}$ & $(2;1)$ & $ 2 \omega_1^{A_1} $ & $2$ \\
\hline
\hline
$3$ & $1$ & ${{U(1)}}$ & $(1;2)$ & $ \omega_1^{A_1} $ & $\frac{1}{2}$ \\
\hline
$4$ & $1$ & ${{A_1}}$ & $(3;0)$ & $ 3 \omega_1^{A_1} $ & $\frac{9}{2} $\\
\hline
\end{tabular}
\caption[smallcaption]{$A_1$ group breaking patterns 
and shift vectors. 
The last component of $n_i'^{(k)}$ corresponds to $n_0'^{(k)}$.}
\label{Tab:A1breaking}
\end{center}
\end{table}

\begin{table}[h!]
\begin{center}
\begin{tabular}{|c|c|c|c|c|c|}
\hline
No. & C.C. & Group breaking & $n_i'^{(k)}$ & Shift vector ($3V$) & $(3V)^2 $ \\
\hline
\hline
$1$ & $0$ & ${{A_2}}$ & $(0,0;3)$ & $ 0 $ & $0 $ \\
\hline
$2$ & $0$ & ${{A_2}}$ & $(0,3;0)$ & $ 3 \omega_2^{A_2} $ & $6 $ \\
\hline
$3$ & $0$ & $U(1)^2$ & $(1,1;1)$ & $  \omega_1^{A_2} +  \omega_2^{A_2} $ & $2$ \\
\hline
$4$ & $0$ & ${{A_2}}$ & $(3,0;0)$ & $ 3 \omega_1^{A_2} $ & $6 $ \\
\hline
\hline
$5$ & $1$ & ${{A_1 \times U(1)}}$ & $(0,2;1)$ & $ 2 \omega_2^{A_2} $ & $\frac{8}{3} $ \\
\hline
$6$ & $1$ & ${{A_1 \times U(1)}}$ & $(1,0;2)$ & $  \omega_1^{A_2} $ & $\frac{2}{3} $ \\
\hline
$7$ & $1$ & ${{A_1 \times U(1)}}$ & $(2,1;0)$ & $ 2 \omega_1^{A_2} +  \omega_2^{A_2} $ & $\frac{14}{3} $ \\
\hline
\end{tabular}
\caption[smallcaption]{$A_2$ group breaking patterns and shift vectors. The last component of $n_i'^{(k)}$ corresponds to $n_0'^{(k)}$. Breaking patterns by shift vectors which belong to conjugacy class $2_{A_2}$ are omitted.}
\label{Tab:A2breaking}
\end{center}
\end{table}

\begin{table}[h!]
\begin{center}
\begin{tabular}{|c|c|c|c|c|c|}
\hline
No. & C.C. & Group breaking & $n_i'^{(k)}$ & Shift vector ($3V$) & $(3V)^2 $ \\
\hline
\hline
$1$ & $0$ & ${{A_4}}$ & $(0,0,0,0;3)$ & $ 0 $ & $0$ \\
\hline
$2$ & $0$ & ${{A_3 \times U(1)}}$ & $(0,0,2,1;0)$ & $ 2 \omega_3^{A_4} + \omega_4^{A_4} $ & $8$ \\
\hline
$3$ & $0$ & ${{A_2 \times A_1 \times U(1)}}$ & $(0,1,0,2;0)$ & $ \omega_2^{A_4} + 2 \omega_4^{A_4} $ & $6$ \\
\hline
$4$ & $0$ & ${{A_1^2 \times U(1)^2}}$ & $(0,1,1,0;1)$ & $ \omega_2^{A_4} + \omega_3^{A_4} $ & $4$ \\
\hline
$5$ & $0$ & ${{A_2 \times U(1)}}$ & $(1,0,0,1;1)$ & $ \omega_1^{A_4} + \omega_4^{A_4} $ & $2$ \\
\hline
$6$ & $0$ & ${{A_3 \times U(1)}}$ & $(1,2,0,0;0)$ & $ \omega_1^{A_4} + 2 \omega_2^{A_4} $ & $8$ \\
\hline
$7$ & $0$ & ${{A_2 \times A_1 \times U(1)}}$ & $(2,0,1,0;0)$ & $ 2 \omega_1^{A_4} + \omega_3^{A_4} $ & $6$ \\
\hline
\hline
$8$ & $1$ & ${{A_3 \times U(1)}}$ & $(0,0,1,2;0)$ & $ \omega_3^{A_4} + 2 \omega_4^{A_4} $ & $\frac{34}{5} $ \\
\hline
$9$ & $1$ & ${{A_2 \times A_1 \times U(1)}}$ & $(0,0,2,0;1)$ & $ 2 \omega_3^{A_4} $ & $\frac{24}{5}$ \\
\hline
$10$ & $1$ & ${{A_1^2 \times U(1)^2}}$ & $(0,1,0,1;1)$ & $ \omega_2^{A_4} + \omega_4^{A_4} $ & $\frac{14}{5} $ \\
\hline
$11$ & $1$ & ${{A_4}}$ & $(0,3,0,0;0)$ & $ 3 \omega_2^{A_4} $ & $\frac{54}{5}$ \\
\hline
$12$ & $1$ & ${{A_3 \times U(1)}}$ & $(1,0,0,0;2)$ & $ \omega_1^{A_4} $ & $\frac{4}{5}$ \\
\hline
$13$ & $1$ & ${{A_2 \times U(1)^2}}$ & $(1,1,1,0;0)$ & $ \omega_1^{A_4} + \omega_2^{A_4} + \omega_3^{A_4}$ & $\frac{34}{5} $ \\
\hline
$14$ & $1$ & ${{A_2 \times A_1 \times U(1)}}$ & $(2,0,0,1;0)$ & $ 2 \omega_1^{A_4} + \omega_4^{A_4} $ & $\frac{24}{5} $ \\
\hline
\hline
$15$ & $2$ & ${{A_4}}$ & $(0,0,0,3;0)$ & $ 3 \omega_4^{A_4}  $ & $\frac{36}{5} $ \\
\hline
$16$ & $2$ & ${{A_2 \times U(1)^2}}$ & $(0,0,1,1;1)$ & $ \omega_3^{A_4} + \omega_4^{A_4} $ & $\frac{16}{5}$ \\
\hline
$17$ & $2$ & ${{A_2 \times A_1 \times U(1)}}$ & $(0,1,0,0;2)$ & $ \omega_2^{A_4} $ & $\frac{6}{5} $ \\
\hline
$18$ & $2$ & ${{A_3 \times U(1)}}$ & $(0,2,1,0;0)$ & $ 2 \omega_2^{A_4} + \omega_3^{A_4} $ & $\frac{46}{5} $ \\
\hline
$19$ & $2$ & ${{A_2 \times A_1 \times U(1)}}$ & $(1,0,2,0;0)$ & $ \omega_1^{A_4} + 2 \omega_3^{A_4} $ & $\frac{36}{5} $ \\
\hline
$20$ & $2$ & ${{A_1^2 \times U(1)^2}}$ & $(1,1,0,1;0)$ & $ \omega_1^{A_4} + \omega_2^{A_4} + \omega_4^{A_4} $ & $\frac{26}{5} $ \\
\hline
$21$ & $2$ & ${{A_3 \times U(1)}}$ & $(2,0,0,0;1)$ & $ 2 \omega_1^{A_4} $ & $\frac{16}{5} $ \\
\hline
\end{tabular}
\caption[smallcaption]{$A_4$ group breaking patterns and shift vectors. The last component of $n_i'^{(k)}$ corresponds to $n_0'^{(k)}$. Breaking patterns by shift vectors which belong to conjugacy classes $3_{A_4}$ and $4_{A_4}$ are omitted.}
\label{Tab:A4breaking}
\end{center}
\end{table}

Next, we consider the ${\bf Z}_3$ asymmetric orbifold model which is specified by the shift vector 
\begin{align}
V = &( 0, \omega_1^{A_1}, 
2 \omega_1^{A_4} + \omega_3^{A_4} - 3 \alpha^{A_4}_1 - 4 \alpha^{A_4}_2 - 2 \alpha^{A_4}_3 - \alpha^{A_4}_4,
- \omega_1^{A_4}  + \alpha^{A_4}_1 + \alpha^{A_4}_2 + \alpha^{A_4}_3 + \alpha^{A_4}_4, \nonumber \\ 
&- \omega_3^{A_4} -2 \omega_4^{A_4} + 2 \alpha^{A_4}_4,
 \omega_2^{A_4} + 2 \omega_4^{A_4} - 2 \alpha^{A_4}_3 - 2 \alpha^{A_4}_4,
\frac{\sqrt{30}}{5},
\frac{3\sqrt{30}}{10},
0, 0,0,0)/3.
\end{align}
This shift vector belongs to the conjugacy class $(0,1,0,4,4,0,1/5,3/10,0,0,0,0)$ of $A_1^2 \times A_4^4 \times U(1)^2 \times A_2 \times \overline{A}_2^3$, i.e. 
\begin{align}
3V \in &( \Gamma_{A_1}, \Gamma_{A_1} + \omega_1^{A_1}, 
\Gamma_{A_4}, \Gamma_{A_4} + \omega_4^{A_4}, \Gamma_{A_4} + \omega_4^{A_4}, \Gamma_{A_4},  \nonumber  \\ 
&\sqrt{30} (n_1 + 1/5),  \sqrt{30} (n_2 + 3/10 ) , \Gamma_{A_2},  
\Gamma_{A_2} , \Gamma_{A_2}, \Gamma_{A_2})
\end{align}
where $\Gamma_{A_1}$, $\Gamma_{A_2}$ and $\Gamma_{A_4}$ are the root lattices of $A_1$, $A_2$ and $A_4$, and $n_1$ and $n_2$ are integers. We can check that the shift vector $V$ satisfies the modular invariance condition \eqref{modinv}. The shift vector breaks the original gauge symmetry $SU(5)^4 \times SU(3) \times SU(2)^2 \times U(1)^2$ to
\begin{align}
SU(4)^2 \times SU(3)^3 \times SU(2)^3 \times U(1)^{7}.
\end{align} 
When we read off the massless spectrum the following description of the shift vector in terms of the unbroken gauge group is useful:
\begin{align}
2 \omega_1^{A_4} + \omega_3^{A_4} - 3 \alpha^{A_4}_1 - 4 \alpha^{A_4}_2 - 2 \alpha^{A_4}_3 - \alpha^{A_4}_4 &= ( - \alpha_1^{A_2} -2 \alpha_2^{A_2},0,0 ), \\
- \omega_1^{A_4}  + \alpha^{A_4}_1 + \alpha^{A_4}_2 + \alpha^{A_4}_3 + \alpha^{A_4}_4 &= ( 0^{A_3}, \frac{2}{\sqrt{5}} ), \\
- \omega_3^{A_4} -2 \omega_4^{A_4} + 2 \alpha^{A_4}_4 &= ( - \omega_3^{A_3}, - \frac{1}{2\sqrt{5}} ), \\
\omega_2^{A_4} + 2 \omega_4^{A_4} - 2 \alpha^{A_4}_3 - 2 \alpha^{A_4}_4 &= ( \alpha_1^{A_2} + 2 \alpha_2^{A_2},0,0 ).
\end{align}

Massless states in the untwisted sector and twisted sector can be read off as in the previous section by using the following information of the lattice. By taking a suitable linear combination of conjugacy class generators, the (20,4)-dimensional part of the invariant sublattice $I_{\theta} ( \Gamma_{20,4} )$ is described as
\begin{align}
&(0,0,3,1,4,2,0,0), \nonumber \\
&(0,1,2,0,3,1,0,1/10), \nonumber \\
&(1,0,1,0,2,2,1/10,0)
\end{align}
of  $ A_1^2 \times A_4^{4} \times U(1)^2 $. 
We can check that this 20-dimensional lattice is spanned by the following basis
\begin{align}
\alpha_{1} &= ( \alpha^{A_1}_{1}, 0,0,0,0,0,0,0 ), \\
\alpha_{2} &= ( 0, \alpha^{A_1}_{1},0,0,0,0,0,0 ), \\
\alpha_{3 \ldots 6} &= ( 0,0,\alpha^{A_4}_{1 \ldots 4}, 0,0,0,0,0 ), \\
\alpha_{7 \ldots 9} &= ( 0,0,0,\alpha^{A_4}_{1 \ldots 3}, 0,0,0,0 ), \\
\alpha_{10 \ldots 13} &= ( 0,0,0,0,\alpha^{A_4}_{1 \ldots 4}, 0,0,0 ), \\
\alpha_{14 \ldots 17} &= ( 0,0,0,0,0,\alpha^{A_4}_{1 \ldots 4}, 0,0 ), \\
\alpha_{18} &= ( 0,0, \omega^{A_4}_3, \omega^{A_4}_1,\omega^{A_4}_4,\omega^{A_4}_2,0,0 ), \\
\alpha_{19} &= ( 0, \omega^{A_4}_1, \omega^{A_4}_2, 0, \omega^{A_4}_3, \omega^{A_4}_1, 0, 1/10 ), \\
\alpha_{20} &= ( \omega^{A_4}_1,0,\omega^{A_4}_1, 0, \omega^{A_4}_2, \omega^{A_4}_2, 1/10, 0 ).
\end{align}
Also, by some calculation we can evaluate a dual basis $\tilde{\alpha}_i$ and obtain that $\tilde{I}_{\theta}( \Gamma_{20,4} )$ is spanned by the conjugacy class generators
\begin{align} 
&( 1,0,0,0,0,0,1/2,0 ), \nonumber \\
&( 0,1,0,0,0,0,0,1/2 ), \nonumber \\
&( 0,0,1,2,0,0,1/5,2/5 ), \nonumber \\
&( 0,0,0,1,1,0,2/5,3/5 ), \nonumber \\
&( 0,0,0,3,0,1,2/5,1/5 ),
\end{align}
with the $U(1)$s normalized to $10/\sqrt{30}$. For the (2,2)-dimensional lattice $\Gamma_{2,2}$, the invariant sublattice $I_{\theta} ( \Gamma_{2,2} )$ is given by the conjugacy class $0$ of $A_2$, so the dual lattice $\tilde{I}_{\theta}( \Gamma_{2,2} )$ is the union of $0$, $1$ and $2$. For this model the degeneracy factor \eqref{Degeneracy} is calculated as
\begin{align}
D= \frac{ \sqrt{3}^3 }{\sqrt{9 \cdot 3}} = 1. 
\end{align}

The resulting massless spectrum of this model is listed in Table \ref{Tab:SpectrumLR1} and \ref{Tab:SpectrumLR2}. The four-dimensional gauge group is $SU(3)_{{\rm C}} \times SU(2)_{{\rm L}} \times SU(2)_{{\rm R}} \times SU(2)_{{\rm F}} \times SU(3)^2 \times SU(4)^2 \times U(1)^7$. This model has $3 \times 5$ fields in the untwisted sector and $65$ fields in the ${\bf Z}_3$ twisted sector. In the table, normalizations for $U(1)^{7}$ charges $U_{1 \ldots 7}$ are taken as 
\begin{align}
U_{1} &= \sqrt{2} Q_{1}, \\
U_{2,5,6,7} &= \sqrt{\frac{5}{6}} Q_{2,5,6,7}, \\
U_{3,4} &= \frac{\sqrt{5}}{2} Q_{3,4}. 
\end{align}
Among the seven $U(1)$ groups, a non-anomalous $U(1)_{{\rm B}-{\rm L}}$ gauge symmetry is taken from a combination of four $U(1)$s,
\begin{align} \label{U1BL}
Q_{{\rm B}-{\rm L}} &= \sqrt{\frac{3}{10}} U_2 -\frac{2}{\sqrt{5}} U_4 -\frac{1}{\sqrt{30}} U_5 + \sqrt{\frac{6}{5}} U_{7}.
\end{align}
There is an anomalous $U(1)_A$ gauge symmetry that is given by the following combination
\begin{align}
Q_A &= \frac{1}{\sqrt{2}} U_1 -\frac{1}{2} \sqrt{\frac{3}{10}} U_2 +\frac{1}{2 \sqrt{5}} U_3 -\frac{1}{2 \sqrt{5}} U_4 + \frac{1}{2} \sqrt{\frac{3}{10}} U_5,
\end{align}
with the GS universality relation by $8 \pi^2 \delta_{GS} = 2 \sqrt{\frac{2}{3}}$.

\begin{table}[h]
\begin{center}
\scriptsize
\begin{tabular}{|c|c|c|c|ccccccc|c|c|c|}
\hline
$U/T$ & $f$ & & ${\rm Irrep.}$ & $Q_1$ & $Q_2$ & $Q_3$ & $Q_4$ & $Q_5$ & $Q_6$ & $Q_7$ 
& $Q_{B-L}$ & $Q_{A}$ & ${{\rm Deg.}}$ \\ 
\hline
${U}$ & $1$ &  & $({\bf 1},{\bf 1},{\bf 1},{\bf 1},{\bf 1},{\bf 1},{\bf 1},{\bf 1})$ & $1$ & $0$ & $0$ & $0$ & $0$ & $0$ & $0$ & $0$ & $1$ & $3$ \\
${U}$ & $2$ &  & $({\bf 1},{\bf 1},{\bf 1},{\bf 1},{\bf 1},{\bf 1},{\bf 4},{\bf 1})$ & $0$ & $0$ & $1$ & $0$ & $0$ & $0$ & $0$ & $0$ & $\frac{1}{4}$ & $3$ \\
${U}$ & $3$ &  & $({\bf 1},{\bf 1},{\bf 1},{\bf 1},{\bf 1},{\bf 1},{\bf 1},\overline{\bf 4})$ & $0$ & $0$ & $0$ & $-1$ & $0$ & $0$ & $0$ & $1$ & $\frac{1}{4}$ & $3$ \\
${U}$ & $4$ &  & $(\overline{\bf 3},{\bf 1},{\bf 2},{\bf 1},{\bf 1},{\bf 1},{\bf 1},{\bf 1})$ & $0$ & $0$ & $0$ & $0$ & $1$ & $0$ & $0$ & $-\frac{1}{6}$ & $\frac{1}{4}$ & $3$ \\
${U}$ & $5$ &  & $({\bf 1},{\bf 2},{\bf 1},{\bf 1},\overline{\bf 3},{\bf 1},{\bf 1},{\bf 1})$ & $0$ & $-1$ & $0$ & $0$ & $0$ & $0$ & $0$ & $-\frac{1}{2}$ & $\frac{1}{4}$ & $3$ \\
${T}$ & $6$ &  & $({\bf 1},{\bf 1},{\bf 1},{\bf 1},{\bf 1},{\bf 1},{\bf 1},{\bf 1})$ & $\frac{1}{6}$ & $-\frac{2}{5}$ & $\frac{4}{15}$ & $\frac{8}{15}$ & $-\frac{4}{5}$ & $\frac{4}{5}$ & $\frac{3}{5}$ & $0$ & $0$ & $1$ \\
${T}$ & $7$ &  & $({\bf 1},{\bf 1},{\bf 1},{\bf 1},{\bf 1},{\bf 1},{\bf 1},{\bf 1})$ & $\frac{1}{6}$ & $\frac{4}{5}$ & $-\frac{8}{15}$ & $-\frac{4}{15}$ & $\frac{2}{5}$ & $-\frac{4}{5}$ & $-\frac{3}{5}$ & $0$ & $0$ & $1$ \\
${T}$ & $8$ &  & $({\bf 1},{\bf 1},{\bf 1},{\bf 1},{\bf 1},{\bf 1},{\bf 1},{\bf 1})$ & $-\frac{1}{3}$ & $-\frac{2}{5}$ & $-\frac{8}{15}$ & $-\frac{4}{15}$ & $-\frac{4}{5}$ & $-\frac{2}{5}$ & $\frac{4}{5}$ & $1$ & $-\frac{1}{2}$ & $1$ \\
${T}$ & $9$ &  & $({\bf 1},{\bf 1},{\bf 1},{\bf 1},{\bf 1},{\bf 1},{\bf 1},{\bf 1})$ & $-\frac{1}{3}$ & $-\frac{2}{5}$ & $\frac{4}{15}$ & $\frac{8}{15}$ & $-\frac{4}{5}$ & $\frac{4}{5}$ & $-\frac{2}{5}$ & $-1$ & $-\frac{1}{2}$ & $1$ \\
${T}$ & $10$ &  & $({\bf 1},{\bf 1},{\bf 1},{\bf 1},{\bf 1},{\bf 1},{\bf 1},{\bf 1})$ & $-\frac{1}{3}$ & $\frac{4}{5}$ & $-\frac{8}{15}$ & $-\frac{4}{15}$ & $\frac{2}{5}$ & $-\frac{4}{5}$ & $\frac{2}{5}$ & $1$ & $-\frac{1}{2}$ & $1$ \\
${T}$ & $11$ &  & $({\bf 1},{\bf 1},{\bf 1},{\bf 1},{\bf 1},{\bf 1},{\bf 1},{\bf 1})$ & $-\frac{1}{3}$ & $\frac{4}{5}$ & $-\frac{8}{15}$ & $\frac{8}{15}$ & $-\frac{4}{5}$ & $0$ & $0$ & $0$ & $-1$ & $1$ \\
${T}$ & $12$ &  & $({\bf 1},{\bf 1},{\bf 1},{\bf 1},{\bf 1},{\bf 1},{\bf 1},{\bf 1})$ & $-\frac{1}{3}$ & $\frac{4}{5}$ & $\frac{4}{15}$ & $\frac{8}{15}$ & $\frac{2}{5}$ & $\frac{2}{5}$ & $-\frac{4}{5}$ & $-1$ & $-\frac{1}{2}$ & $1$ \\
${T}$ & $13$ &  & $({\bf 1},{\bf 1},{\bf 2},{\bf 2},{\bf 3},{\bf 1},{\bf 1},{\bf 1})$ & $\frac{1}{6}$ & $0$ & $\frac{4}{15}$ & $-\frac{4}{15}$ & $-\frac{1}{5}$ & $\frac{1}{5}$ & $\frac{1}{5}$ & $\frac{1}{2}$ & $\frac{1}{4}$ & $1$ \\
${T}$ & $14$ &  & $({\bf 1},{\bf 1},{\bf 2},{\bf 1},{\bf 3},{\bf 1},{\bf 1},{\bf 1})$ & $\frac{1}{6}$ & $0$ & $\frac{4}{15}$ & $-\frac{4}{15}$ & $-\frac{1}{5}$ & $-\frac{4}{5}$ & $\frac{1}{5}$ & $\frac{1}{2}$ & $\frac{1}{4}$ & $1$ \\
${T}$ & $15$ &  & $({\bf 1},{\bf 1},{\bf 2},{\bf 1},{\bf 1},{\bf 1},{\bf 4},{\bf 1})$ & $\frac{1}{6}$ & $-\frac{2}{5}$ & $-\frac{1}{3}$ & $\frac{8}{15}$ & $-\frac{1}{5}$ & $0$ & $\frac{1}{5}$ & $-\frac{1}{2}$ & $0$ & $1$ \\
${T}$ & $16$ &  & $({\bf 1},{\bf 1},{\bf 2},{\bf 1},{\bf 1},{\bf 1},\overline{\bf 4},{\bf 1})$ & $\frac{1}{6}$ & $\frac{4}{5}$ & $\frac{1}{15}$ & $-\frac{4}{15}$ & $-\frac{1}{5}$ & $0$ & $-\frac{1}{5}$ & $\frac{1}{2}$ & $0$ & $1$ \\
${T}$ & $17$ &  & $({\bf 1},{\bf 1},{\bf 2},{\bf 1},{\bf 1},{\bf 1},{\bf 1},{\bf 6})$ & $\frac{1}{6}$ & $-\frac{2}{5}$ & $\frac{4}{15}$ & $\frac{2}{15}$ & $-\frac{1}{5}$ & $\frac{2}{5}$ & $-\frac{1}{5}$ & $-\frac{1}{2}$ & $\frac{1}{4}$ & $1$ \\
${T}$ & $18$ &  & $({\bf 1},{\bf 1},{\bf 1},{\bf 2},{\bf 1},{\bf 1},{\bf 1},{\bf 1})$ & $\frac{1}{6}$ & $-\frac{2}{5}$ & $-\frac{8}{15}$ & $-\frac{4}{15}$ & $-\frac{4}{5}$ & $\frac{3}{5}$ & $-\frac{1}{5}$ & $0$ & $0$ & $1$ \\
${T}$ & $19$ &  & $({\bf 1},{\bf 1},{\bf 1},{\bf 2},{\bf 1},{\bf 1},{\bf 1},{\bf 1})$ & $\frac{1}{6}$ & $-\frac{2}{5}$ & $\frac{4}{15}$ & $\frac{8}{15}$ & $-\frac{4}{5}$ & $-\frac{1}{5}$ & $\frac{3}{5}$ & $0$ & $0$ & $1$ \\
${T}$ & $20$ &  & $({\bf 1},{\bf 1},{\bf 1},{\bf 2},{\bf 1},{\bf 1},{\bf 1},{\bf 1})$ & $\frac{1}{6}$ & $\frac{4}{5}$ & $-\frac{8}{15}$ & $-\frac{4}{15}$ & $\frac{2}{5}$ & $\frac{1}{5}$ & $-\frac{3}{5}$ & $0$ & $0$ & $1$ \\
${T}$ & $21$ &  & $({\bf 1},{\bf 1},{\bf 1},{\bf 2},{\bf 1},{\bf 1},{\bf 1},{\bf 1})$ & $\frac{1}{6}$ & $\frac{4}{5}$ & $\frac{4}{15}$ & $\frac{8}{15}$ & $\frac{2}{5}$ & $-\frac{3}{5}$ & $\frac{1}{5}$ & $0$ & $0$ & $1$ \\
${T}$ & $22$ &  & $({\bf 1},{\bf 1},{\bf 1},{\bf 2},{\bf 1},{\bf 1},{\bf 1},{\bf 1})$ & $-\frac{1}{3}$ & $-\frac{2}{5}$ & $\frac{4}{15}$ & $-\frac{4}{15}$ & $\frac{2}{5}$ & $-1$ & $0$ & $0$ & $0$ & $1$ \\
${T}$ & $23$ &  & $({\bf 1},{\bf 1},{\bf 1},{\bf 2},{\bf 1},{\bf 1},{\bf 1},{\bf 1})$ & $-\frac{1}{3}$ & $-\frac{2}{5}$ & $\frac{4}{15}$ & $-\frac{4}{15}$ & $\frac{2}{5}$ & $1$ & $0$ & $0$ & $0$ & $1$ \\
${T}$ & $24$ &  & $({\bf 1},{\bf 1},{\bf 1},{\bf 2},{\bf 1},{\bf 1},{\bf 1},{\bf 1})$ & $-\frac{1}{3}$ & $-\frac{2}{5}$ & $\frac{4}{15}$ & $\frac{8}{15}$ & $-\frac{4}{5}$ & $-\frac{1}{5}$ & $-\frac{2}{5}$ & $-1$ & $-\frac{1}{2}$ & $1$ \\
${T}$ & $25$ &  & $({\bf 1},{\bf 1},{\bf 1},{\bf 2},{\bf 1},{\bf 1},{\bf 1},{\bf 1})$ & $-\frac{1}{3}$ & $\frac{4}{5}$ & $-\frac{8}{15}$ & $-\frac{4}{15}$ & $\frac{2}{5}$ & $\frac{1}{5}$ & $\frac{2}{5}$ & $1$ & $-\frac{1}{2}$ & $1$ \\
${T}$ & $26$ &  & $({\bf 1},{\bf 1},{\bf 1},{\bf 2},{\bf 1},{\bf 1},\overline{\bf 4},{\bf 1})$ & $\frac{1}{6}$ & $-\frac{2}{5}$ & $\frac{1}{15}$ & $\frac{8}{15}$ & $\frac{2}{5}$ & $\frac{1}{5}$ & $-\frac{1}{5}$ & $-1$ & $\frac{1}{4}$ & $1$ \\
${T}$ & $27$ &  & $({\bf 1},{\bf 1},{\bf 1},{\bf 2},{\bf 1},{\bf 1},{\bf 1},{\bf 4})$ & $\frac{1}{6}$ & $-\frac{2}{5}$ & $-\frac{8}{15}$ & $-\frac{1}{15}$ & $\frac{2}{5}$ & $-\frac{1}{5}$ & $\frac{1}{5}$ & $0$ & $\frac{1}{4}$ & $1$ \\
${T}$ & $28$ &  & $({\bf 1},{\bf 1},{\bf 1},{\bf 1},{\bf 3},{\bf 1},{\bf 1},{\bf 1})$ & $\frac{1}{6}$ & $0$ & $-\frac{8}{15}$ & $\frac{8}{15}$ & $\frac{2}{5}$ & $-\frac{2}{5}$ & $\frac{3}{5}$ & $0$ & $0$ & $1$ \\
${T}$ & $29$ &  & $({\bf 1},{\bf 1},{\bf 1},{\bf 1},{\bf 3},{\bf 1},{\bf 1},{\bf 1})$ & $-\frac{1}{3}$ & $0$ & $-\frac{8}{15}$ & $\frac{8}{15}$ & $\frac{2}{5}$ & $-\frac{2}{5}$ & $-\frac{2}{5}$ & $-1$ & $-\frac{1}{2}$ & $1$ \\
${T}$ & $30$ &  & $({\bf 1},{\bf 1},{\bf 1},{\bf 1},{\bf 3},{\bf 1},{\bf 4},{\bf 1})$ & $\frac{1}{6}$ & $0$ & $-\frac{1}{3}$ & $-\frac{4}{15}$ & $\frac{2}{5}$ & $\frac{2}{5}$ & $-\frac{1}{5}$ & $0$ & $\frac{1}{4}$ & $1$ \\
${T}$ & $31$ &  & $({\bf 1},{\bf 1},{\bf 1},{\bf 1},\overline{\bf 3},{\bf 1},{\bf 1},{\bf 1})$ & $\frac{1}{6}$ & $\frac{2}{5}$ & $\frac{4}{15}$ & $-\frac{4}{15}$ & $-\frac{4}{5}$ & $\frac{2}{5}$ & $-\frac{3}{5}$ & $0$ & $0$ & $1$ \\
${T}$ & $32$ &  & $({\bf 1},{\bf 1},{\bf 1},{\bf 1},\overline{\bf 3},{\bf 1},{\bf 1},{\bf 1})$ & $-\frac{1}{3}$ & $\frac{2}{5}$ & $\frac{4}{15}$ & $-\frac{4}{15}$ & $-\frac{4}{5}$ & $\frac{2}{5}$ & $\frac{2}{5}$ & $1$ & $-\frac{1}{2}$ & $1$ \\
${T}$ & $33$ &  & $({\bf 1},{\bf 1},{\bf 1},{\bf 1},\overline{\bf 3},{\bf 1},{\bf 1},{\bf 4})$ & $\frac{1}{6}$ & $\frac{2}{5}$ & $\frac{4}{15}$ & $-\frac{1}{15}$ & $\frac{2}{5}$ & $-\frac{2}{5}$ & $-\frac{1}{5}$ & $0$ & $\frac{1}{4}$ & $1$ \\
${T}$ & $34$ &  & $({\bf 1},{\bf 1},{\bf 1},{\bf 1},{\bf 1},{\bf 3},{\bf 1},{\bf 1})$ & $\frac{1}{6}$ & $-\frac{2}{5}$ & $-\frac{8}{15}$ & $-\frac{4}{15}$ & $-\frac{4}{5}$ & $-\frac{2}{5}$ & $-\frac{1}{5}$ & $0$ & $0$ & $1$ \\
${T}$ & $35$ &  & $({\bf 1},{\bf 1},{\bf 1},{\bf 1},{\bf 1},{\bf 3},{\bf 1},{\bf 1})$ & $\frac{1}{6}$ & $-\frac{2}{5}$ & $\frac{4}{15}$ & $-\frac{4}{15}$ & $\frac{2}{5}$ & $0$ & $-1$ & $-1$ & $\frac{1}{2}$ & $1$ \\
${T}$ & $36$ &  & $({\bf 1},{\bf 1},{\bf 1},{\bf 1},{\bf 1},{\bf 3},{\bf 1},{\bf 1})$ & $\frac{1}{6}$ & $-\frac{2}{5}$ & $\frac{4}{15}$ & $-\frac{4}{15}$ & $\frac{2}{5}$ & $0$ & $1$ & $1$ & $\frac{1}{2}$ & $1$ \\
${T}$ & $37$ &  & $({\bf 1},{\bf 1},{\bf 1},{\bf 1},{\bf 1},{\bf 3},{\bf 1},{\bf 1})$ & $\frac{1}{6}$ & $\frac{4}{5}$ & $\frac{4}{15}$ & $\frac{8}{15}$ & $\frac{2}{5}$ & $\frac{2}{5}$ & $\frac{1}{5}$ & $0$ & $0$ & $1$ \\
${T}$ & $38$ &  & $({\bf 1},{\bf 1},{\bf 1},{\bf 1},{\bf 1},{\bf 3},{\bf 1},{\bf 1})$ & $\frac{2}{3}$ & $-\frac{2}{5}$ & $\frac{4}{15}$ & $-\frac{4}{15}$ & $\frac{2}{5}$ & $0$ & $0$ & $0$ & $1$ & $1$ \\
${T}$ & $39$ &  & $({\bf 1},{\bf 1},{\bf 1},{\bf 1},{\bf 1},\overline{\bf 3},{\bf 1},{\bf 1})$ & $\frac{1}{6}$ & $-\frac{2}{5}$ & $-\frac{8}{15}$ & $-\frac{4}{15}$ & $-\frac{4}{5}$ & $-\frac{2}{5}$ & $-\frac{1}{5}$ & $0$ & $0$ & $1$ \\
${T}$ & $40$ &  & $({\bf 1},{\bf 1},{\bf 1},{\bf 1},{\bf 1},\overline{\bf 3},{\bf 1},{\bf 1})$ & $\frac{1}{6}$ & $-\frac{2}{5}$ & $\frac{4}{15}$ & $-\frac{4}{15}$ & $\frac{2}{5}$ & $0$ & $-1$ & $-1$ & $\frac{1}{2}$ & $1$ \\
${T}$ & $41$ &  & $({\bf 1},{\bf 1},{\bf 1},{\bf 1},{\bf 1},\overline{\bf 3},{\bf 1},{\bf 1})$ & $\frac{1}{6}$ & $-\frac{2}{5}$ & $\frac{4}{15}$ & $-\frac{4}{15}$ & $\frac{2}{5}$ & $0$ & $1$ & $1$ & $\frac{1}{2}$ & $1$ \\
${T}$ & $42$ &  & $({\bf 1},{\bf 1},{\bf 1},{\bf 1},{\bf 1},\overline{\bf 3},{\bf 1},{\bf 1})$ & $\frac{1}{6}$ & $\frac{4}{5}$ & $\frac{4}{15}$ & $\frac{8}{15}$ & $\frac{2}{5}$ & $\frac{2}{5}$ & $\frac{1}{5}$ & $0$ & $0$ & $1$ \\
${T}$ & $43$ &  & $({\bf 1},{\bf 1},{\bf 1},{\bf 1},{\bf 1},\overline{\bf 3},{\bf 1},{\bf 1})$ & $\frac{2}{3}$ & $-\frac{2}{5}$ & $\frac{4}{15}$ & $-\frac{4}{15}$ & $\frac{2}{5}$ & $0$ & $0$ & $0$ & $1$ & $1$ \\
${T}$ & $44$ &  & $({\bf 1},{\bf 1},{\bf 1},{\bf 1},{\bf 1},{\bf 1},{\bf 4},{\bf 1})$ & $\frac{1}{6}$ & $-\frac{2}{5}$ & $\frac{7}{15}$ & $-\frac{4}{15}$ & $-\frac{4}{5}$ & $-\frac{2}{5}$ & $-\frac{1}{5}$ & $0$ & $\frac{1}{4}$ & $1$ \\
${T}$ & $45$ &  & $({\bf 1},{\bf 1},{\bf 1},{\bf 1},{\bf 1},{\bf 1},\overline{\bf 4},{\bf 1})$ & $\frac{1}{6}$ & $-\frac{2}{5}$ & $\frac{1}{15}$ & $\frac{8}{15}$ & $\frac{2}{5}$ & $-\frac{4}{5}$ & $-\frac{1}{5}$ & $-1$ & $\frac{1}{4}$ & $1$ \\
${T}$ & $46$ &  & $({\bf 1},{\bf 1},{\bf 1},{\bf 1},{\bf 1},{\bf 1},\overline{\bf 4},{\bf 1})$ & $-\frac{1}{3}$ & $-\frac{2}{5}$ & $-\frac{11}{15}$ & $-\frac{4}{15}$ & $\frac{2}{5}$ & $0$ & $0$ & $0$ & $-\frac{1}{4}$ & $1$ \\
${T}$ & $47$ &  & $({\bf 1},{\bf 1},{\bf 1},{\bf 1},{\bf 1},{\bf 1},{\bf 1},{\bf 4})$ & $\frac{1}{6}$ & $-\frac{2}{5}$ & $-\frac{8}{15}$ & $-\frac{1}{15}$ & $\frac{2}{5}$ & $\frac{4}{5}$ & $\frac{1}{5}$ & $0$ & $\frac{1}{4}$ & $1$ \\
${T}$ & $48$ &  & $({\bf 1},{\bf 1},{\bf 1},{\bf 1},{\bf 1},{\bf 1},{\bf 1},{\bf 4})$ & $-\frac{1}{3}$ & $-\frac{2}{5}$ & $\frac{4}{15}$ & $\frac{11}{15}$ & $\frac{2}{5}$ & $0$ & $0$ & $-1$ & $-\frac{1}{4}$ & $1$ \\
${T}$ & $49$ &  & $({\bf 1},{\bf 1},{\bf 1},{\bf 1},{\bf 1},{\bf 1},{\bf 1},\overline{\bf 4})$ & $\frac{1}{6}$ & $\frac{4}{5}$ & $\frac{4}{15}$ & $-\frac{7}{15}$ & $\frac{2}{5}$ & $\frac{2}{5}$ & $\frac{1}{5}$ & $1$ & $\frac{1}{4}$ & $1$ \\
${T}$ & $50$ &  & $({\bf 3},{\bf 1},{\bf 1},{\bf 1},{\bf 1},{\bf 1},{\bf 1},{\bf 1})$ & $\frac{1}{6}$ & $-\frac{2}{5}$ & $-\frac{8}{15}$ & $\frac{8}{15}$ & $0$ & $\frac{2}{5}$ & $-\frac{3}{5}$ & $-\frac{4}{3}$ & $0$ & $1$ \\
${T}$ & $51$ &  & $({\bf 3},{\bf 1},{\bf 1},{\bf 1},{\bf 1},{\bf 1},{\bf 1},{\bf 1})$ & $-\frac{1}{3}$ & $-\frac{2}{5}$ & $-\frac{8}{15}$ & $\frac{8}{15}$ & $0$ & $\frac{2}{5}$ & $\frac{2}{5}$ & $-\frac{1}{3}$ & $-\frac{1}{2}$ & $1$ \\
${T}$ & $52$ &  & $({\bf 3},{\bf 1},{\bf 2},{\bf 1},{\bf 1},{\bf 1},{\bf 1},{\bf 1})$ & $-\frac{1}{3}$ & $-\frac{2}{5}$ & $\frac{4}{15}$ & $-\frac{4}{15}$ & $-\frac{3}{5}$ & $0$ & $0$ & $\frac{1}{6}$ & $-\frac{1}{4}$ & $1$ \\
${T}$ & $53$ &  & $({\bf 3},{\bf 1},{\bf 1},{\bf 1},\overline{\bf 3},{\bf 1},{\bf 1},{\bf 1})$ & $\frac{1}{6}$ & $\frac{2}{5}$ & $-\frac{8}{15}$ & $-\frac{4}{15}$ & $0$ & $0$ & $\frac{1}{5}$ & $\frac{2}{3}$ & $0$ & $1$ \\
${T}$ & $54$ &  & $({\bf 3},{\bf 1},{\bf 1},{\bf 1},{\bf 1},{\bf 1},{\bf 1},\overline{\bf 4})$ & $\frac{1}{6}$ & $-\frac{2}{5}$ & $\frac{4}{15}$ & $\frac{1}{3}$ & $0$ & $-\frac{2}{5}$ & $\frac{1}{5}$ & $-\frac{1}{3}$ & $\frac{1}{4}$ & $1$ \\
\hline
\end{tabular}
\caption[smallcaption]{Massless spectrum of three-generation $SU(3)_{{\rm C}} \times SU(2)_{{\rm L}} \times SU(2)_{{\rm R}} \times U(1)_{{\rm B}-{\rm L}}$ model. Representations under the non-Abelian group $SU(3)_{{\rm C}} \times SU(2)_{{\rm L}} \times SU(2)_{{\rm R}} \times SU(2)_{{\rm F}} \times SU(3)^2 \times SU(4)^2$ and $U(1)$ charges are listed. U and T mean the untwisted and twisted sector respectively. Note that the degeneracy of untwisted fields is 3, while it is 1 for twisted fields. The gravity and gauge supermultiplets are omitted.}
\label{Tab:SpectrumLR1}
\end{center}
\end{table}

\begin{table}[h]
\begin{center}
\scriptsize
\begin{tabular}{|c|c|c|c|ccccccc|c|c|c|}
\hline
$U/T$ & $f$ & & ${\rm Irrep.}$ & $Q_1$ & $Q_2$ & $Q_3$ & $Q_4$ & $Q_5$ & $Q_6$ & $Q_7$ 
& $Q_{B-L}$ & $Q_{A}$ & ${{\rm Deg.}}$ \\ 
\hline
${T}$ & $55$ &  & $(\overline{\bf 3},{\bf 1},{\bf 1},{\bf 1},{\bf 1},{\bf 1},{\bf 1},{\bf 1})$ & $\frac{1}{6}$ & $\frac{4}{5}$ & $\frac{4}{15}$ & $-\frac{4}{15}$ & $-\frac{2}{5}$ & $-\frac{2}{5}$ & $\frac{3}{5}$ & $\frac{4}{3}$ & $0$ & $1$ \\
${T}$ & $56$ &  & $(\overline{\bf 3},{\bf 1},{\bf 1},{\bf 1},{\bf 1},{\bf 1},{\bf 1},{\bf 1})$ & $-\frac{1}{3}$ & $\frac{4}{5}$ & $\frac{4}{15}$ & $-\frac{4}{15}$ & $-\frac{2}{5}$ & $-\frac{2}{5}$ & $-\frac{2}{5}$ & $\frac{1}{3}$ & $-\frac{1}{2}$ & $1$ \\
${T}$ & $57$ & $Q_{{\rm R}3}$ & $(\overline{\bf 3},{\bf 1},{\bf 2},{\bf 1},{\bf 1},{\bf 1},{\bf 1},{\bf 1})$ & $\frac{1}{6}$ & $-\frac{2}{5}$ & $-\frac{8}{15}$ & $-\frac{4}{15}$ & $\frac{1}{5}$ & $-\frac{2}{5}$ & $-\frac{1}{5}$ & $-\frac{1}{6}$ & $\frac{1}{4}$ & $1$ \\
${T}$ & $58$ &  & $(\overline{\bf 3},{\bf 1},{\bf 1},{\bf 1},{\bf 3},{\bf 1},{\bf 1},{\bf 1})$ & $\frac{1}{6}$ & $0$ & $\frac{4}{15}$ & $\frac{8}{15}$ & $-\frac{2}{5}$ & $0$ & $-\frac{1}{5}$ & $-\frac{2}{3}$ & $0$ & $1$ \\
${T}$ & $59$ &  & $(\overline{\bf 3},{\bf 1},{\bf 1},{\bf 1},{\bf 1},{\bf 1},\overline{\bf 4},{\bf 1})$ & $\frac{1}{6}$ & $-\frac{2}{5}$ & $\frac{1}{15}$ & $-\frac{4}{15}$ & $-\frac{2}{5}$ & $\frac{2}{5}$ & $\frac{1}{5}$ & $\frac{1}{3}$ & $\frac{1}{4}$ & $1$ \\
${T}$ & $60$ &  & $({\bf 1},{\bf 2},{\bf 2},{\bf 1},{\bf 1},{\bf 1},{\bf 1},{\bf 1})$ & $\frac{1}{6}$ & $\frac{1}{5}$ & $-\frac{8}{15}$ & $-\frac{4}{15}$ & $-\frac{1}{5}$ & $\frac{2}{5}$ & $\frac{3}{5}$ & $1$ & $0$ & $1$ \\
${T}$ & $61$ &  & $({\bf 1},{\bf 2},{\bf 2},{\bf 1},{\bf 1},{\bf 1},{\bf 1},{\bf 1})$ & $\frac{1}{6}$ & $\frac{1}{5}$ & $\frac{4}{15}$ & $\frac{8}{15}$ & $-\frac{1}{5}$ & $-\frac{2}{5}$ & $-\frac{3}{5}$ & $-1$ & $0$ & $1$ \\
${T}$ & $62$ &  & $({\bf 1},{\bf 2},{\bf 2},{\bf 1},{\bf 1},{\bf 1},{\bf 1},{\bf 1})$ & $-\frac{1}{3}$ & $\frac{1}{5}$ & $-\frac{8}{15}$ & $-\frac{4}{15}$ & $-\frac{1}{5}$ & $\frac{2}{5}$ & $-\frac{2}{5}$ & $0$ & $-\frac{1}{2}$ & $1$ \\
${T}$ & $63$ & $H$ & $({\bf 1},{\bf 2},{\bf 2},{\bf 1},{\bf 1},{\bf 1},{\bf 1},{\bf 1})$ & $-\frac{1}{3}$ & $\frac{1}{5}$ & $\frac{4}{15}$ & $\frac{8}{15}$ & $-\frac{1}{5}$ & $-\frac{2}{5}$ & $\frac{2}{5}$ & $0$ & $-\frac{1}{2}$ & $1$ \\
${T}$ & $64$ &  & $({\bf 1},{\bf 2},{\bf 1},{\bf 1},{\bf 3},{\bf 1},{\bf 1},{\bf 1})$ & $-\frac{1}{3}$ & $\frac{3}{5}$ & $\frac{4}{15}$ & $-\frac{4}{15}$ & $\frac{2}{5}$ & $0$ & $0$ & $\frac{1}{2}$ & $-\frac{1}{4}$ & $1$ \\
${T}$ & $65$ &  & $({\bf 1},{\bf 2},{\bf 1},{\bf 1},\overline{\bf 3},{\bf 1},{\bf 1},{\bf 1})$ & $\frac{1}{6}$ & $-\frac{1}{5}$ & $\frac{4}{15}$ & $\frac{8}{15}$ & $\frac{2}{5}$ & $\frac{2}{5}$ & $\frac{1}{5}$ & $-\frac{1}{2}$ & $\frac{1}{4}$ & $1$ \\
${T}$ & $66$ &  & $({\bf 1},{\bf 2},{\bf 1},{\bf 1},{\bf 1},{\bf 1},{\bf 6},{\bf 1})$ & $\frac{1}{6}$ & $\frac{1}{5}$ & $-\frac{2}{15}$ & $-\frac{4}{15}$ & $\frac{2}{5}$ & $-\frac{2}{5}$ & $\frac{1}{5}$ & $\frac{1}{2}$ & $\frac{1}{4}$ & $1$ \\
${T}$ & $67$ &  & $({\bf 1},{\bf 2},{\bf 1},{\bf 1},{\bf 1},{\bf 1},{\bf 1},{\bf 4})$ & $\frac{1}{6}$ & $\frac{1}{5}$ & $\frac{4}{15}$ & $-\frac{1}{15}$ & $-\frac{4}{5}$ & $0$ & $\frac{1}{5}$ & $\frac{1}{2}$ & $0$ & $1$ \\
${T}$ & $68$ &  & $({\bf 1},{\bf 2},{\bf 1},{\bf 1},{\bf 1},{\bf 1},{\bf 1},\overline{\bf 4})$ & $\frac{1}{6}$ & $\frac{1}{5}$ & $-\frac{8}{15}$ & $\frac{1}{3}$ & $\frac{2}{5}$ & $0$ & $-\frac{1}{5}$ & $-\frac{1}{2}$ & $0$ & $1$ \\
${T}$ & $69$ & $Q_{{\rm L}{\bf 2}}$ & $({\bf 3},{\bf 2},{\bf 1},{\bf 2},{\bf 1},{\bf 1},{\bf 1},{\bf 1})$ & $\frac{1}{6}$ & $\frac{1}{5}$ & $\frac{4}{15}$ & $-\frac{4}{15}$ & $0$ & $-\frac{1}{5}$ & $-\frac{1}{5}$ & $\frac{1}{6}$ & $\frac{1}{4}$ & $1$ \\
${T}$ & $70$ & $Q_{{\rm L}{\bf 1}}$ & $({\bf 3},{\bf 2},{\bf 1},{\bf 1},{\bf 1},{\bf 1},{\bf 1},{\bf 1})$ & $\frac{1}{6}$ & $\frac{1}{5}$ & $\frac{4}{15}$ & $-\frac{4}{15}$ & $0$ & $\frac{4}{5}$ & $-\frac{1}{5}$ & $\frac{1}{6}$ & $\frac{1}{4}$ & $1$ \\
\hline
\end{tabular}
\caption[smallcaption]{Massless spectrum of three-generation $SU(3)_{{\rm C}} \times SU(2)_{{\rm L}} \times SU(2)_{{\rm R}} \times U(1)_{{\rm B}-{\rm L}}$ model (continued). Representations under the non-Abelian group $SU(3)_{{\rm C}} \times SU(2)_{{\rm L}} \times SU(2)_{{\rm R}} \times SU(2)_{{\rm F}} \times SU(3)^2 \times SU(4)^2$ and $U(1)$ charges are listed. U and T mean the untwisted and twisted sector respectively. Note that the degeneracy of untwisted fields is 3, while it is 1 for twisted fields. The gravity and gauge supermultiplets are omitted.}
\label{Tab:SpectrumLR2}
\end{center}
\end{table}

By choosing this combination, we can see that this model contains three chiral matter generations of a supersymmetric left-right symmetric model, and the additional fields have vector-like structure. This model has ten singlets of the non-Abelian part $SU(3)_{{\rm C}} \times SU(2)_{{\rm L}} \times SU(2)_{{\rm R}} \times SU(2)_{{\rm F}} \times SU(3)^2 \times SU(4)^2$. The number of fields in this model is relatively small, $3 \times 5 + 65 = 80$. The fields $f_{69}$ and $f_{70}$ in the ${\bf Z}_3$ twisted sector are identified as three generations of left-handed quarks $Q_{{\rm L}}$ $( {\bf 3}, {\bf 2}, {\bf 1} )_{1/6}$, referred to as $Q_{{\rm L}{\bf 2}}$ and $Q_{{\rm L}{\bf 1}}$. Interestingly, this model has a $SU(2)_{{\rm F}}$ gauge flavor symmetry which unifies the first two generations of $Q_{{\rm L}}$ fields into the $SU(2)_{{\rm F}}$ doublet $Q_{{\rm L}{\bf 2}}$, and the third generation as a $SU(2)_{{\rm F}}$ singlet $Q_{{\rm L}{\bf 1}}$. The possibility to have flavor gauge symmetries is a characteristic property of the asymmetric orbifold construction: since we do not consider any left-moving twist action, the zero point energy for the left-movers does not increase. Then, states in a non-trivial representation of a flavor gauge symmetry can be realized at massless level. Fields $f_{18 \ldots 23}$ are $SU(2)_{{\rm F}}$ doublets without B-L charge, so these fields can be candidates for a $SU(2)_{{\rm F}}$ flavon field. This model does not contain matter fields in the adjoint representation. Therefore, we expect $SU(2)_{{\rm R}} \times U(1)_{{\rm B}-{\rm L}}$ to be broken by a VEV of the doublet fields $( {\bf 2}, +1/2 )$ and $( {\bf 2}, -1/2 )$. 

Regarding the top quark mass, a 3-point coupling for the top Yukawa interaction $H Q_{{\rm L}{\bf 1}}  Q_{{\rm R}3}$ is not allowed by the Q-charge invariance though this operator is gauge invariant. Then, to reproduce a suitable value for the top quark mass, we need to consider higher-dimensional operators and larger VEVs for some singlets.

\section{Conclusion}

From ${\bf Z}_3$ heterotic asymmetric orbifolds, we construct several three-generation models with the standard model gauge group or the left-right symmetric group. The starting points for model building are Narain lattices with $\overline{E}_6$ or $\overline{A}_2^3$ which are obtained from 24-dimensional Niemeier lattices by the lattice engineering technique. By taking a modular invariant combination of ${\bf Z}_3$ shift actions for the left-mover we obtain four-dimensional three-generation models with vector-like exotics. 

For models from Narain lattices with $\overline{E}_6$, the number of "three" standard model generations originates from the degeneracy factor $D=3$. Also, the up-type quark resides in the ${\bf Z}_3$ twisted sector and its Yukawa coupling is expected to be realized through a coupling of three fields from the twisted sector. By Q-charge analysis it turns out that, for one of the models, there is a three point coupling for top Yukawa interaction. However, the other three-point couplings lead to a too heavy mass for the second generation quark. Regarding the model with degeneracy factor $D=1$, even though this is a left-right symmetric model with three generations, charge conservation forbids a three-point coupling for top Yukawa interaction. So, in order to reproduce appropriate Yukawa couplings we will need to search models from other Narain lattices with $\overline{A}_2^3$. 

In asymmetric orbifold constructions, it turns out that it is possible to construct models with a gauge flavor symmetry. It will be important to consider Yukawa interaction properties (masses and mixings) arising from such kind of flavor gauge symmetries as well as those arising from discrete flavor symmetries. Analyzing moduli stabilization in this formalism will also be important.


\subsection*{Acknowledgement}
F.B. was supported by the "Leadership Development Program for Space Exploration and Research" from the Japan Society for the Promotion of Science. T.K. was supported in part by the Grant-in-Aid for Scientific Research No.~25400252 from the Ministry of Education, Culture, Sports, Science and Technology of Japan. S.K. was supported by the Taiwan's National Science Council under grants NSC102-2811-M-033-002 and NSC102-2811-M-033-008.

\appendix


\section{Another example of a three-generation model} \label{App1}

Here, we show another example of a simple three-generation model with the standard model group. To specify the model, we take the (22,6)-dimensional Narain lattice $A_2^{11} \times \overline{E}_6$ as a starting point, and also take a ${\bf Z}_3$ shift vector as 
\begin{align}
V= ( \omega_2^{A_2}, \omega_2^{A_2}, 2 \omega_2^{A_2}, 0, 
\omega_2^{A_2}, \omega_2^{A_2}, \omega_2^{A_2}, \omega_2^{A_2}, 
2 \omega_2^{A_2}, - \omega_2^{A_2}, \alpha_1^{A_2}, 0  )/3.
\end{align}
This shift vector belongs to the conjugacy class $(2, 2, 1, 0, 2, 2, 2, 2, 1, 1, 0 ,0)$ of $A_2^{11} \times \overline{E}_6$. By the orbifold action, the original gauge group $SU(3)^{11}$ breaks to
\begin{align}
SU(3) \times SU(2)^{9} \times U(1)^{11},
\end{align}
and chiral supermultiplets of this model are summarized in Table \ref{Tab:A211model}. This model is a three-generation model with $SU(3)_{{\rm C}} \times SU(2)_{{\rm L}} \times U(1)$ group. The fields non-trivially charged under $SU(3)_{{\rm C}} \times SU(2)_{{\rm L}}$ are
\begin{align}
3\left\{
( {\bf 3}, {\bf 2} ),
 2 ( \overline{{\bf 3}}, {\bf 1} ),
 5 ({\bf 1}, {\bf 2} )
 \right\}.
\end{align}
Also, there are $3 \times 10$ singlets under the non-Abelian group $SU(3)_{{\rm C}} \times SU(2)_{{\rm L}} \times SU(2)^{8}$, the other fields are charged under the hidden group $SU(2)^8$ as the (bi-)fundamental representation. It turns out that this model has no color exotic fields, and the number of extra lepton doublet fields is very small ($3 \times 2$). However, by gauge invariance there is no three point interaction for an up-type Yukawa coupling in this model. Then, to reproduce a suitable value for the top quark mass, we have to take into account higher-dimensional operators and larger VEVs for some singlets.

\begin{table}[h]
\begin{center}
\scriptsize
\begin{tabular}{|c|c|c|ccccccccccc|c|}
\hline
$U/T$ & $f$ & ${\rm Irrep.}$ & $Q_1$ & $Q_2$ & $Q_3$ & $Q_4$ & $Q_5$ & $Q_6$ & $Q_7$ & $Q_8$ & $Q_9$ & $Q_{10}$ & $Q_{11}$ & ${{\rm Deg.}}$ 
 \\ 
\hline
 U & 1 & $( {\bf 1},{\bf 1};{\bf 1},{\bf 1},{\bf 1},{\bf 1},{\bf 1},{\bf 1},{\bf 1},{\bf 1} )$ & $0$ & $0$ & $0$ & $0$ & $0$ & $0$ & $0$ & $0$ & $0$ & $\frac{1}{2}$ & $-1$ & $3$ \\
 U & 2 & $( {\bf 1},{\bf 1};{\bf 1},{\bf 1},{\bf 1},{\bf 1},{\bf 1},{\bf 1},{\bf 1},{\bf 1} )$ & $0$ & $0$ & $0$ & $0$ & $0$ & $0$ & $0$ & $0$ & $0$ & $\frac{1}{2}$ & $1$ & $3$ \\
 U & 3 & $( {\bf 1},{\bf 1};{\bf 1},{\bf 1},{\bf 1},{\bf 1},{\bf 1},{\bf 1},{\bf 1},{\bf 1} )$ & $0$ & $0$ & $0$ & $0$ & $0$ & $0$ & $0$ & $0$ & $0$ & $-1$ & $0$ & $3$ \\
 U & 4 & $( {\bf 1},{\bf 1};{\bf 2},{\bf 1},{\bf 1},{\bf 1},{\bf 1},{\bf 1},{\bf 1},{\bf 1} )$ & $1$ & $0$ & $0$ & $0$ & $0$ & $0$ & $0$ & $0$ & $0$ & $0$ & $0$ & $3$ \\
 U & 5 & $( {\bf 1},{\bf 1};{\bf 1},{\bf 2},{\bf 1},{\bf 1},{\bf 1},{\bf 1},{\bf 1},{\bf 1} )$ & $0$ & $1$ & $0$ & $0$ & $0$ & $0$ & $0$ & $0$ & $0$ & $0$ & $0$ & $3$ \\
 U & 6 & $( {\bf 1},{\bf 1};{\bf 1},{\bf 1},{\bf 2},{\bf 1},{\bf 1},{\bf 1},{\bf 1},{\bf 1} )$ & $0$ & $0$ & $-1$ & $0$ & $0$ & $0$ & $0$ & $0$ & $0$ & $0$ & $0$ & $3$ \\
 U & 7 & $( {\bf 1},{\bf 1};{\bf 1},{\bf 1},{\bf 1},{\bf 2},{\bf 1},{\bf 1},{\bf 1},{\bf 1} )$ & $0$ & $0$ & $0$ & $1$ & $0$ & $0$ & $0$ & $0$ & $0$ & $0$ & $0$ & $3$ \\
 U & 8 & $( {\bf 1},{\bf 1};{\bf 1},{\bf 1},{\bf 1},{\bf 1},{\bf 2},{\bf 1},{\bf 1},{\bf 1} )$ & $0$ & $0$ & $0$ & $0$ & $1$ & $0$ & $0$ & $0$ & $0$ & $0$ & $0$ & $3$ \\
 U & 9 & $( {\bf 1},{\bf 1};{\bf 1},{\bf 1},{\bf 1},{\bf 1},{\bf 1},{\bf 2},{\bf 1},{\bf 1} )$ & $0$ & $0$ & $0$ & $0$ & $0$ & $1$ & $0$ & $0$ & $0$ & $0$ & $0$ & $3$ \\
 U & 10 & $( {\bf 1},{\bf 1};{\bf 1},{\bf 1},{\bf 1},{\bf 1},{\bf 1},{\bf 1},{\bf 2},{\bf 1} )$ & $0$ & $0$ & $0$ & $0$ & $0$ & $0$ & $1$ & $0$ & $0$ & $0$ & $0$ & $3$ \\
 U & 11 & $( {\bf 1},{\bf 1};{\bf 1},{\bf 1},{\bf 1},{\bf 1},{\bf 1},{\bf 1},{\bf 1},{\bf 2} )$ & $0$ & $0$ & $0$ & $0$ & $0$ & $0$ & $0$ & $-1$ & $0$ & $0$ & $0$ & $3$ \\
 U & 12 & $( {\bf 1},{\bf 2};{\bf 1},{\bf 1},{\bf 1},{\bf 1},{\bf 1},{\bf 1},{\bf 1},{\bf 1} )$ & $0$ & $0$ & $0$ & $0$ & $0$ & $0$ & $0$ & $0$ & $-1$ & $0$ & $0$ & $3$ \\
 T & 13 & $({\bf 1},{\bf 1};{\bf 1},{\bf 1},{\bf 1},{\bf 1},{\bf 1},{\bf 1},{\bf 1},{\bf 1} )$ 
& $\frac{2}{9}$ & $\frac{2}{9}$ & $\frac{4}{9}$ & $\frac{2}{9}$ & $\frac{2}{9}$ & $\frac{2}{9}$ & $\frac{2}{9}$ & $\frac{4}{9}$ & $-\frac{2}{9}$ & $-\frac{2}{3}$ & $0$ & $3$ \\
 T & 14 & $({\bf 1},{\bf 1};{\bf 1},{\bf 1},{\bf 1},{\bf 1},{\bf 1},{\bf 1},{\bf 1},{\bf 1} )$ 
& $-\frac{4}{9}$ & $-\frac{4}{9}$ & $-\frac{2}{9}$ & $-\frac{4}{9}$ & $\frac{2}{9}$ & $\frac{2}{9}$ & $-\frac{4}{9}$ & $\frac{4}{9}$ & $-\frac{2}{9}$ & $\frac{1}{3}$ & $0$ & $3$ \\
 T & 15 & $({\bf 1},{\bf 1};{\bf 1},{\bf 1},{\bf 1},{\bf 1},{\bf 1},{\bf 1},{\bf 1},{\bf 1} )$ 
& $\frac{2}{9}$ & $\frac{2}{9}$ & $-\frac{2}{9}$ & $\frac{2}{9}$ & $\frac{2}{9}$ & $-\frac{4}{9}$ & $-\frac{4}{9}$ & $-\frac{2}{9}$ & $-\frac{2}{9}$ & $\frac{1}{3}$ & $-\frac{2}{3}$ & $3$ \\
 T & 16 & $({\bf 1},{\bf 1};{\bf 1},{\bf 1},{\bf 1},{\bf 1},{\bf 1},{\bf 1},{\bf 1},{\bf 1} )$ 
& $-\frac{4}{9}$ & $-\frac{4}{9}$ & $-\frac{2}{9}$ & $-\frac{4}{9}$ & $-\frac{4}{9}$ & $\frac{2}{9}$ & $\frac{2}{9}$ & $-\frac{2}{9}$ & $\frac{4}{9}$ & $-\frac{1}{6}$ & $-\frac{1}{3}$ & $3$ \\
 T & 17 & $({\bf 1},{\bf 1};{\bf 1},{\bf 1},{\bf 1},{\bf 1},{\bf 1},{\bf 1},{\bf 1},{\bf 1} )$ 
& $-\frac{4}{9}$ & $-\frac{4}{9}$ & $\frac{4}{9}$ & $\frac{2}{9}$ & $-\frac{4}{9}$ & $\frac{2}{9}$ & $-\frac{4}{9}$ & $-\frac{2}{9}$ & $-\frac{2}{9}$ & $-\frac{1}{6}$ & $\frac{1}{3}$ & $3$ \\
 T & 18 & $({\bf 1},{\bf 1};{\bf 1},{\bf 1},{\bf 1},{\bf 1},{\bf 1},{\bf 1},{\bf 1},{\bf 1} )$ 
& $\frac{2}{9}$ & $-\frac{4}{9}$ & $-\frac{2}{9}$ & $-\frac{4}{9}$ & $-\frac{4}{9}$ & $-\frac{4}{9}$ & $\frac{2}{9}$ & $\frac{4}{9}$ & $-\frac{2}{9}$ & $-\frac{1}{6}$ & $\frac{1}{3}$ & $3$ \\
 T & 19 & $({\bf 1},{\bf 1};{\bf 1},{\bf 1},{\bf 1},{\bf 1},{\bf 1},{\bf 1},{\bf 1},{\bf 1} )$ 
& $-\frac{4}{9}$ & $\frac{2}{9}$ & $\frac{4}{9}$ & $-\frac{4}{9}$ & $-\frac{4}{9}$ & $-\frac{4}{9}$ & $\frac{2}{9}$ & $-\frac{2}{9}$ & $-\frac{2}{9}$ & $\frac{1}{3}$ & $0$ & $3$ \\
 T & 20 & $({\bf 1},{\bf 1};{\bf 2},{\bf 1},{\bf 1},{\bf 1},{\bf 1},{\bf 1},{\bf 1},{\bf 1} )$ 
& $-\frac{1}{9}$ & $\frac{2}{9}$ & $\frac{4}{9}$ & $-\frac{4}{9}$ & $\frac{2}{9}$ & $\frac{2}{9}$ & $-\frac{4}{9}$ & $-\frac{2}{9}$ & $-\frac{2}{9}$ & $-\frac{1}{6}$ & $-\frac{1}{3}$ & $3$ \\
 T & 21 & $({\bf 1},{\bf 1};{\bf 1},{\bf 2},{\bf 1},{\bf 1},{\bf 1},{\bf 1},{\bf 1},{\bf 1} )$ 
& $\frac{2}{9}$ & $-\frac{1}{9}$ & $-\frac{2}{9}$ & $\frac{2}{9}$ & $-\frac{4}{9}$ & $\frac{2}{9}$ & $-\frac{4}{9}$ & $\frac{4}{9}$ & $-\frac{2}{9}$ & $-\frac{1}{6}$ & $-\frac{1}{3}$ & $3$ \\
 T & 22 & $({\bf 1},{\bf 1};{\bf 1},{\bf 1},{\bf 2},{\bf 1},{\bf 1},{\bf 1},{\bf 1},{\bf 1} )$ 
& $\frac{2}{9}$ & $-\frac{4}{9}$ & $\frac{1}{9}$ & $\frac{2}{9}$ & $\frac{2}{9}$ & $\frac{2}{9}$ & $-\frac{4}{9}$ & $-\frac{2}{9}$ & $\frac{4}{9}$ & $\frac{1}{3}$ & $0$ & $3$ \\
 T & 23 & $({\bf 1},{\bf 1};{\bf 1},{\bf 1},{\bf 2},{\bf 1},{\bf 1},{\bf 1},{\bf 1},{\bf 1} )$ 
& $\frac{2}{9}$ & $\frac{2}{9}$ & $-\frac{5}{9}$ & $\frac{2}{9}$ & $\frac{2}{9}$ & $\frac{2}{9}$ & $\frac{2}{9}$ & $\frac{4}{9}$ & $-\frac{2}{9}$ & $\frac{1}{3}$ & $0$ & $3$ \\
 T & 24 & $({\bf 1},{\bf 1};{\bf 1},{\bf 1},{\bf 2},{\bf 1},{\bf 1},{\bf 1},{\bf 1},{\bf 1} )$ 
& $\frac{2}{9}$ & $\frac{2}{9}$ & $\frac{1}{9}$ & $-\frac{4}{9}$ & $\frac{2}{9}$ & $-\frac{4}{9}$ & $\frac{2}{9}$ & $-\frac{2}{9}$ & $\frac{4}{9}$ & $-\frac{1}{6}$ & $-\frac{1}{3}$ & $3$ \\
 T & 25 & $({\bf 1},{\bf 1};{\bf 1},{\bf 1},{\bf 1},{\bf 2},{\bf 1},{\bf 1},{\bf 1},{\bf 1} )$ 
& $-\frac{4}{9}$ & $\frac{2}{9}$ & $-\frac{2}{9}$ & $-\frac{1}{9}$ & $\frac{2}{9}$ & $-\frac{4}{9}$ & $\frac{2}{9}$ & $\frac{4}{9}$ & $-\frac{2}{9}$ & $-\frac{1}{6}$ & $-\frac{1}{3}$ & $3$ \\
 T & 26 & $({\bf 1},{\bf 1};{\bf 1},{\bf 1},{\bf 1},{\bf 1},{\bf 2},{\bf 1},{\bf 1},{\bf 1} )$ 
& $\frac{2}{9}$ & $-\frac{4}{9}$ & $\frac{4}{9}$ & $\frac{2}{9}$ & $-\frac{1}{9}$ & $-\frac{4}{9}$ & $\frac{2}{9}$ & $-\frac{2}{9}$ & $-\frac{2}{9}$ & $-\frac{1}{6}$ & $-\frac{1}{3}$ & $3$ \\
 T & 27 & $({\bf 1},{\bf 1};{\bf 1},{\bf 1},{\bf 1},{\bf 1},{\bf 1},{\bf 2},{\bf 1},{\bf 1} )$ 
& $\frac{2}{9}$ & $\frac{2}{9}$ & $-\frac{2}{9}$ & $\frac{2}{9}$ & $\frac{2}{9}$ & $\frac{5}{9}$ & $-\frac{4}{9}$ & $-\frac{2}{9}$ & $-\frac{2}{9}$ & $-\frac{1}{6}$ & $\frac{1}{3}$ & $3$ \\
 T & 28 & $({\bf 1},{\bf 1};{\bf 1},{\bf 1},{\bf 1},{\bf 1},{\bf 1},{\bf 2},{\bf 1},{\bf 1} )$ 
& $\frac{2}{9}$ & $-\frac{4}{9}$ & $-\frac{2}{9}$ & $\frac{2}{9}$ & $\frac{2}{9}$ & $-\frac{1}{9}$ & $\frac{2}{9}$ & $\frac{4}{9}$ & $\frac{4}{9}$ & $-\frac{1}{6}$ & $-\frac{1}{3}$ & $3$ \\
 T & 29 & $({\bf 1},{\bf 1};{\bf 1},{\bf 1},{\bf 1},{\bf 1},{\bf 1},{\bf 2},{\bf 1},{\bf 1} )$ 
& $-\frac{4}{9}$ & $\frac{2}{9}$ & $\frac{4}{9}$ & $\frac{2}{9}$ & $\frac{2}{9}$ & $-\frac{1}{9}$ & $\frac{2}{9}$ & $-\frac{2}{9}$ & $\frac{4}{9}$ & $-\frac{1}{6}$ & $\frac{1}{3}$ & $3$ \\
 T & 30 & $({\bf 1},{\bf 1};{\bf 1},{\bf 1},{\bf 1},{\bf 1},{\bf 1},{\bf 1},{\bf 2},{\bf 1} )$ 
& $\frac{2}{9}$ & $\frac{2}{9}$ & $-\frac{2}{9}$ & $-\frac{4}{9}$ & $\frac{2}{9}$ & $\frac{2}{9}$ & $-\frac{1}{9}$ & $\frac{4}{9}$ & $\frac{4}{9}$ & $-\frac{1}{6}$ & $\frac{1}{3}$ & $3$ \\
 T & 31 & $({\bf 1},{\bf 1};{\bf 1},{\bf 1},{\bf 1},{\bf 1},{\bf 1},{\bf 1},{\bf 2},{\bf 1} )$ 
& $\frac{2}{9}$ & $\frac{2}{9}$ & $-\frac{2}{9}$ & $\frac{2}{9}$ & $\frac{2}{9}$ & $-\frac{4}{9}$ & $\frac{5}{9}$ & $-\frac{2}{9}$ & $-\frac{2}{9}$ & $-\frac{1}{6}$ & $\frac{1}{3}$ & $3$ \\
 T & 32 & $({\bf 1},{\bf 1};{\bf 1},{\bf 1},{\bf 1},{\bf 1},{\bf 1},{\bf 1},{\bf 2},{\bf 1} )$ 
& $\frac{2}{9}$ & $\frac{2}{9}$ & $\frac{4}{9}$ & $\frac{2}{9}$ & $-\frac{4}{9}$ & $\frac{2}{9}$ & $-\frac{1}{9}$ & $-\frac{2}{9}$ & $\frac{4}{9}$ & $-\frac{1}{6}$ & $-\frac{1}{3}$ & $3$ \\
 T & 33 & $({\bf 1},{\bf 1};{\bf 1},{\bf 1},{\bf 1},{\bf 1},{\bf 1},{\bf 1},{\bf 1},{\bf 2} )$ 
& $\frac{2}{9}$ & $\frac{2}{9}$ & $\frac{4}{9}$ & $\frac{2}{9}$ & $\frac{2}{9}$ & $\frac{2}{9}$ & $\frac{2}{9}$ & $-\frac{5}{9}$ & $-\frac{2}{9}$ & $\frac{1}{3}$ & $0$ & $3$ \\
 T & 34 & $({\bf 1},{\bf 1};{\bf 1},{\bf 1},{\bf 1},{\bf 1},{\bf 1},{\bf 1},{\bf 1},{\bf 2} )$ 
& $-\frac{4}{9}$ & $\frac{2}{9}$ & $-\frac{2}{9}$ & $\frac{2}{9}$ & $\frac{2}{9}$ & $\frac{2}{9}$ & $-\frac{4}{9}$ & $\frac{1}{9}$ & $\frac{4}{9}$ & $-\frac{1}{6}$ & $-\frac{1}{3}$ & $3$ \\
 T & 35 & $({\bf 1},{\bf 1};{\bf 1},{\bf 1},{\bf 1},{\bf 1},{\bf 1},{\bf 1},{\bf 1},{\bf 2} )$ 
& $\frac{2}{9}$ & $\frac{2}{9}$ & $-\frac{2}{9}$ & $\frac{2}{9}$ & $-\frac{4}{9}$ & $-\frac{4}{9}$ & $\frac{2}{9}$ & $\frac{1}{9}$ & $\frac{4}{9}$ & $\frac{1}{3}$ & $0$ & $3$ \\
 T & 36 & $({\bf 1},{\bf 1};{\bf 2},{\bf 1},{\bf 1},{\bf 2},{\bf 1},{\bf 1},{\bf 1},{\bf 1} )$ 
& $-\frac{1}{9}$ & $\frac{2}{9}$ & $-\frac{2}{9}$ & $-\frac{1}{9}$ & $\frac{2}{9}$ & $\frac{2}{9}$ & $\frac{2}{9}$ & $-\frac{2}{9}$ & $\frac{4}{9}$ & $\frac{1}{3}$ & $0$ & $3$ \\
 T & 37 & $({\bf 1},{\bf 1};{\bf 2},{\bf 1},{\bf 1},{\bf 1},{\bf 1},{\bf 2},{\bf 1},{\bf 1} )$ 
& $-\frac{1}{9}$ & $\frac{2}{9}$ & $-\frac{2}{9}$ & $\frac{2}{9}$ & $-\frac{4}{9}$ & $-\frac{1}{9}$ & $\frac{2}{9}$ & $-\frac{2}{9}$ & $-\frac{2}{9}$ & $-\frac{1}{6}$ & $-\frac{1}{3}$ & $3$ \\
 T & 38 & $({\bf 1},{\bf 1};{\bf 2},{\bf 1},{\bf 1},{\bf 1},{\bf 1},{\bf 1},{\bf 1},{\bf 2} )$ 
& $-\frac{1}{9}$ & $-\frac{4}{9}$ & $-\frac{2}{9}$ & $\frac{2}{9}$ & $\frac{2}{9}$ & $\frac{2}{9}$ & $\frac{2}{9}$ & $\frac{1}{9}$ & $-\frac{2}{9}$ & $-\frac{1}{6}$ & $\frac{1}{3}$ & $3$ \\
 T & 39 & $({\bf 1},{\bf 1};{\bf 1},{\bf 2},{\bf 2},{\bf 1},{\bf 1},{\bf 1},{\bf 1},{\bf 1} )$ 
& $-\frac{4}{9}$ & $-\frac{1}{9}$ & $\frac{1}{9}$ & $\frac{2}{9}$ & $\frac{2}{9}$ & $\frac{2}{9}$ & $\frac{2}{9}$ & $-\frac{2}{9}$ & $-\frac{2}{9}$ & $-\frac{1}{6}$ & $-\frac{1}{3}$ & $3$ \\
 T & 40 & $({\bf 1},{\bf 1};{\bf 1},{\bf 2},{\bf 1},{\bf 1},{\bf 2},{\bf 1},{\bf 1},{\bf 1} )$ 
& $\frac{2}{9}$ & $-\frac{1}{9}$ & $-\frac{2}{9}$ & $\frac{2}{9}$ & $-\frac{1}{9}$ & $\frac{2}{9}$ & $\frac{2}{9}$ & $-\frac{2}{9}$ & $\frac{4}{9}$ & $-\frac{1}{6}$ & $\frac{1}{3}$ & $3$ \\
 T & 41 & $({\bf 1},{\bf 1};{\bf 1},{\bf 2},{\bf 1},{\bf 1},{\bf 1},{\bf 2},{\bf 1},{\bf 1} )$ 
& $\frac{2}{9}$ & $-\frac{1}{9}$ & $-\frac{2}{9}$ & $-\frac{4}{9}$ & $\frac{2}{9}$ & $-\frac{1}{9}$ & $\frac{2}{9}$ & $-\frac{2}{9}$ & $-\frac{2}{9}$ & $\frac{1}{3}$ & $0$ & $3$ \\
 T & 42 & $({\bf 1},{\bf 1};{\bf 1},{\bf 1},{\bf 2},{\bf 2},{\bf 1},{\bf 1},{\bf 1},{\bf 1} )$ 
& $\frac{2}{9}$ & $\frac{2}{9}$ & $\frac{1}{9}$ & $-\frac{1}{9}$ & $-\frac{4}{9}$ & $\frac{2}{9}$ & $\frac{2}{9}$ & $-\frac{2}{9}$ & $-\frac{2}{9}$ & $-\frac{1}{6}$ & $\frac{1}{3}$ & $3$ \\
 T & 43 & $({\bf 1},{\bf 1};{\bf 1},{\bf 1},{\bf 1},{\bf 2},{\bf 1},{\bf 1},{\bf 2},{\bf 1} )$ 
& $\frac{2}{9}$ & $-\frac{4}{9}$ & $-\frac{2}{9}$ & $-\frac{1}{9}$ & $\frac{2}{9}$ & $\frac{2}{9}$ & $-\frac{1}{9}$ & $-\frac{2}{9}$ & $-\frac{2}{9}$ & $-\frac{1}{6}$ & $-\frac{1}{3}$ & $3$ \\
 T & 44 & $({\bf 1},{\bf 1};{\bf 1},{\bf 1},{\bf 1},{\bf 1},{\bf 2},{\bf 1},{\bf 2},{\bf 1} )$ 
& $-\frac{4}{9}$ & $\frac{2}{9}$ & $-\frac{2}{9}$ & $\frac{2}{9}$ & $-\frac{1}{9}$ & $\frac{2}{9}$ & $-\frac{1}{9}$ & $-\frac{2}{9}$ & $-\frac{2}{9}$ & $\frac{1}{3}$ & $0$ & $3$ \\
 T & 45 & $({\bf 1},{\bf 1};{\bf 1},{\bf 1},{\bf 1},{\bf 1},{\bf 2},{\bf 1},{\bf 1},{\bf 2} )$ 
& $\frac{2}{9}$ & $\frac{2}{9}$ & $-\frac{2}{9}$ & $-\frac{4}{9}$ & $-\frac{1}{9}$ & $\frac{2}{9}$ & $\frac{2}{9}$ & $\frac{1}{9}$ & $-\frac{2}{9}$ & $-\frac{1}{6}$ & $-\frac{1}{3}$ & $3$ \\
 T & 46 & $({\bf 1},{\bf 2};{\bf 1},{\bf 1},{\bf 1},{\bf 1},{\bf 1},{\bf 1},{\bf 1},{\bf 1} )$ 
& $\frac{2}{9}$ & $-\frac{4}{9}$ & $\frac{4}{9}$ & $-\frac{4}{9}$ & $\frac{2}{9}$ & $\frac{2}{9}$ & $\frac{2}{9}$ & $-\frac{2}{9}$ & $\frac{1}{9}$ & $-\frac{1}{6}$ & $\frac{1}{3}$ & $3$ \\
 T & 47 & $({\bf 1},{\bf 2};{\bf 1},{\bf 1},{\bf 1},{\bf 1},{\bf 1},{\bf 1},{\bf 1},{\bf 1} )$ 
& $-\frac{4}{9}$ & $-\frac{4}{9}$ & $-\frac{2}{9}$ & $\frac{2}{9}$ & $\frac{2}{9}$ & $-\frac{4}{9}$ & $\frac{2}{9}$ & $-\frac{2}{9}$ & $\frac{1}{9}$ & $\frac{1}{3}$ & $0$ & $3$ \\
 T & 48 & $({\bf 1},{\bf 2};{\bf 1},{\bf 1},{\bf 1},{\bf 1},{\bf 1},{\bf 1},{\bf 1},{\bf 1} )$ 
& $-\frac{4}{9}$ & $\frac{2}{9}$ & $-\frac{2}{9}$ & $\frac{2}{9}$ & $-\frac{4}{9}$ & $\frac{2}{9}$ & $\frac{2}{9}$ & $\frac{4}{9}$ & $\frac{1}{9}$ & $-\frac{1}{6}$ & $\frac{1}{3}$ & $3$ \\
 T & 49 & $({\bf 1},{\bf 2};{\bf 1},{\bf 1},{\bf 1},{\bf 1},{\bf 1},{\bf 1},{\bf 1},{\bf 1} )$ 
& $\frac{2}{9}$ & $\frac{2}{9}$ & $-\frac{2}{9}$ & $-\frac{4}{9}$ & $-\frac{4}{9}$ & $\frac{2}{9}$ & $-\frac{4}{9}$ & $-\frac{2}{9}$ & $\frac{1}{9}$ & $\frac{1}{3}$ & $0$ & $3$ \\
 T & 50 & $(\overline{{\bf 3}},{\bf 1};{\bf 1},{\bf 1},{\bf 1},{\bf 1},{\bf 1},{\bf 1},{\bf 1},{\bf 1} )$ 
& $\frac{2}{9}$ & $-\frac{4}{9}$ & $-\frac{2}{9}$ & $\frac{2}{9}$ & $-\frac{4}{9}$ & $\frac{2}{9}$ & $\frac{2}{9}$ & $-\frac{2}{9}$ & $-\frac{2}{9}$ & $\frac{1}{3}$ & $0$ & $3$ \\
 T & 51 & $(\overline{{\bf 3}},{\bf 1};{\bf 1},{\bf 1},{\bf 1},{\bf 1},{\bf 1},{\bf 1},{\bf 1},{\bf 1} )$ 
& $-\frac{4}{9}$ & $\frac{2}{9}$ & $-\frac{2}{9}$ & $-\frac{4}{9}$ & $\frac{2}{9}$ & $\frac{2}{9}$ & $\frac{2}{9}$ & $-\frac{2}{9}$ & $-\frac{2}{9}$ & $-\frac{1}{6}$ & $\frac{1}{3}$ & $3$ \\
 T & 52 & $({\bf 3},{\bf 2};{\bf 1},{\bf 1},{\bf 1},{\bf 1},{\bf 1},{\bf 1},{\bf 1},{\bf 1} )$ 
& $\frac{2}{9}$ & $\frac{2}{9}$ & $-\frac{2}{9}$ & $\frac{2}{9}$ & $\frac{2}{9}$ & $\frac{2}{9}$ & $\frac{2}{9}$ & $-\frac{2}{9}$ & $\frac{1}{9}$ & $-\frac{1}{6}$ & $-\frac{1}{3}$ & $3$ \\
\hline
\end{tabular}
\caption[smallcaption]{Massless spectrum of three-generation $SU(3)_{{\rm C}} \times SU(2)_{{\rm L}} \times U(1)$ model. Representations under the non-Abelian group $SU(3)_{{\rm C}} \times SU(2)_{{\rm L}} \times SU(2)^8 $ and $U(1)$ charges are listed. U and T mean the untwisted and twisted sector respectively. Note that all fields have degeneracy 3. The gravity and gauge supermultiplets are omitted.}
\label{Tab:A211model}
\end{center}
\end{table}

\end{document}